\documentclass[%
 aip,
 amsmath,amssymb,
 reprint,%
]{revtex4-1}

\usepackage{graphicx}
\usepackage{dcolumn}
\usepackage{bm}

\usepackage[utf8]{inputenc}
\usepackage[T1]{fontenc}
\usepackage{mathptmx}
\usepackage{newtxtext}

\usepackage[dvipsnames,usenames]{color}

\newcommand{\llrrangle}[1]{
  \left\langle\mkern-3mu\left\langle#1\right\rangle\mkern-3mu\right\rangle}

\begin{document}

\preprint{AIP/123-QED}

\title{Probing anharmonic phonons by quantum correlators: A path integral approach}

\author{T. Morresi}
\affiliation{Institut de Minéralogie, de Physique des Matériaux et de Cosmochimie (IMPMC), Sorbonne Université, CNRS UMR 7590, IRD UMR 206, MNHN, 4 Place Jussieu, 75252 Paris, France}

\author{L. Paulatto}
\affiliation{Institut de Minéralogie, de Physique des Matériaux et de Cosmochimie (IMPMC), Sorbonne Université, CNRS UMR 7590, IRD UMR 206, MNHN, 4 Place Jussieu, 75252 Paris, France}
\author{R. Vuilleumier}%
\affiliation{PASTEUR, Département de chimie, École normale supérieure, PSL University, Sorbonne Université, CNRS, 75005 Paris, France}
\author{M. Casula}
\email{michele.casula@sorbonne-universite.fr}
\affiliation{Institut de Minéralogie, de Physique des Matériaux et de Cosmochimie (IMPMC), Sorbonne Université, CNRS UMR 7590, IRD UMR 206, MNHN, 4 Place Jussieu, 75252 Paris, France}


\begin{abstract}
We devise an efficient scheme 
to determine vibrational properties  
from Path Integral Molecular Dynamics (PIMD) simulations. The method is based on 
zero-time Kubo-transformed correlation functions and 
captures the anharmonicity of the potential due to both temperature and quantum effects. Using analytical derivations and numerical calculations on toy-model potentials,
we show that two different estimators built upon PIMD correlation functions 
fully characterize the phonon spectra and the anharmonicity strength. 
The first estimator is associated with force-force quantum correlators and gives access to the fundamental frequencies and thermodynamic properties of the quantum system. The second one is instead connected to displacement-displacement correlators and probes the lowest-energy phonon excitations with high accuracy. We also prove that the use of generalized eigenvalue equations, in place of the standard normal mode equations,
leads to a significant speed-up in the PIMD phonon calculations, 
both in terms of faster convergence rate and smaller time-step bias. 
Within this framework, using ab initio PIMD simulations, we compute phonon dispersions of diamond and of the high-pressure I4$_1$/amd phase of atomic hydrogen. We find that, in the latter case, the anharmonicity is stronger than previously estimated and 
yields a sizeable red-shift in
the vibrational spectrum of 
atomic hydrogen.
\end{abstract}

\maketitle

%

\section{Introduction}
\label{intro}
Atomic vibrations are responsible for many thermodynamic 
and spectral properties of molecules and solids \cite{Born_1954, Wilson_1955}, such as the thermal expansion and specific heat, or Raman and infrared spectra, to mention just a few. Furthermore, since the advent of the
Bardeen, Cooper, and Schrieffer (BCS) theory, \cite{Bardeen_1957}
it is well 
established
that 
lattice vibrations
are fundamental to determine the superconducting materials properties 
through their coupling with electrons \cite{Stewart_2107}. In particular, the recent exciting discoveries of high-temperature high-pressure hydrogen-based superconductors, such as H$_3$S  \cite{Drozdov_2015} and LaH$_{10}$ \cite{Drozdov_2019}, have revealed the key role played by hydrogen, the lightest element of the periodic table, in making these materials superconducting. The distinguishing feature of these systems is the presence of strong proton fluctuations
and large
electron-phonon coupling, which both drive the system into the superconducting state. However, while the impact of nuclear quantum effects (NQE) on the behaviour of hydrogen-based materials is 
substantial,
from the theoretical and computational sides the inclusion of NQE 
into phonon dispersion still represents a challenge for strongly anharmonic systems.

The standard framework to get lattice and ionic vibrational properties is the harmonic approximation, where the static Born-Oppenheimer (BO) potential is expanded until the second order around the potential minimum, that is the equilibrium configuration of the system. While this approach is the basis for 
beyond-harmonic theories, it neglects the eventual anharmonicity due to 
NQE and temperature.
For instance, one cannot explain the thermal expansion of solids \cite{Fultz_2006} within this framework.
In order to go beyond the harmonic 
approximation
and take into account 
interactions between phonons, 
one can attempt to use perturbation theory \cite{Paulatto_2013}. 
This is however doomed to failure whenever applied to 
anharmonic systems such as hydrogen-rich materials, because in these cases
large atomic fluctuations go well beyond the range 
of applicability
required by 
theory. 
To overcome this difficulty, one can use methods based on Molecular Dynamics (MD) simulations, where anharmonic effects can be taken into account non-perturbatively. MD methods have been applied to molecules \cite{Schmitz_2004,Martinez_2006} to compute normal modes, and to crystal systems \cite{Hellman_2011,Hellman_2013,Koukaras_2015,Kong_2011} to compute phonons in a variety of ways \footnote{Throughout this work we refer to phonons for both molecules and crystal systems}. 
Nevertheless,
classical MD simulations can describe thermal effects, but they do not include the quantum statistical behaviours of nuclei. 
The quantum thermal bath \cite{Dammak_2009,Dammak_2012} has been introduced in this regard
to mimic NQE in MD simulations. This scheme is based on a modified Langevin dynamics and is exact in the case of systems made of harmonic oscillators. It provides satisfactory results in some anharmonic systems such as ice \cite{Bronstein_2014,Bronstein_2016} but, being based on classical MD, it suffers from zero point energy leakage issues \cite{Ben-Nun_1996}, although some recent work tends to correct this problem \cite{Mangaud_2019}. 
Methods not based on MD while including both thermal and quantum 
effects are the Self Consistent Harmonic Approximation \cite{Hooton_1955} (SCHA) and its stochastic implementations \cite{Errea_2013,Bianco_2017}(SSCHA), and the Self Consistent Ab Initio Lattice Dynamics \cite{Souvatzis_2008} (SCAILD), where anharmonicity is sampled through real space stochastic techniques.
Also the Self Consistent Phonon (SCPH) theory, developed in Ref.~\onlinecite{Werthamer_1970} and implemented in Ref.~\onlinecite{Tadano_2015}, belongs to this class of non-MD schemes.

Despite their recent revival, mainly boosted by the physics of hydrogen-rich materials, all these methods deteriorate in case of strong anharmonicity.
Path integral techniques are a way to improve upon previous self-consistent approaches and to go towards sampling the \emph{exact} thermal distribution of quantum nuclei, therefore including anharmonicity at all orders. However, the price to pay is the increase in computational cost and the difficulty of \emph{measuring} the vibrational properties, once the quantum thermal distribution is sampled.

In this work, we propose a scheme to extract accurate phonon dispersions and low-lying excitations from the quantum thermal distribution function 
exactly 
sampled by efficient Path Integral Molecular Dynamics (PIMD) simulations. While PIMD is a general framework that covers many types of path integrals combined into 
MD
simulations, such as Centroid Molecular Dynamics, Ring Polymer Molecular Dynamics (RPMD) or its thermostatted version (TRPMD), in the following we will always refer to PIMD 
in the context
of TRPMD. In our approach, the quantum system is thermalized through a Langevin thermostat, implemented within the framework of the 
fast
Path Integral Ornstein-Uhlenbeck Dynamics (PIOUD) algorithm \cite{Mouhat_2017}.

By extending the maximum localization criterion to determine normal modes in classical MD \cite{Martinez_2006}, we develop quantum Kubo-transformed correlation functions that give rise to analogous expressions in PIMD. We prove that these quantum correlators are one order of magnitude more efficient than the standard ones, by yielding converged phonon frequencies after a few picoseconds of dynamics and at a larger time steps. We first test the proposed estimators in one- and two-dimensional toy models, which are exactly solvable. Then, we apply our estimators to compute the phonon dispersion of materials from first principles, by using forces from density functional theory (DFT). We show the case of diamond, a benchmark system, and the more difficult atomic I4$_1$/amd phase of hydrogen, where NQE and anharmonic effects are large.

The paper is organized as follows: In Sec.~\ref{sec:theory} we
set the formalism of our
PIMD framework,
and we define the quantum correlators 
upon which the 
phonon estimators are based, valid for both isolated and extended systems.
Results are shown in Sec.~\ref{sec:results}. In particular, in Sec.~\ref{sec:results_1d} we focus on one-dimensional (1D) toy models; in Sec.~\ref{sec:results_2d} we study a strongly anharmonic two-dimensional (2D) system; in Sec.~\ref{sec:results_diamond} we report \emph{ab initio} results obtained for diamond and in Sec.~\ref{sec:results_hydrogen} the ones for atomic I4$_1$/amd hydrogen. Finally, we draw conclusions and perspectives in Sec.~\ref{sec:conclusion}.  

\section{Theory}
\label{sec:theory}
We present here the PIMD framework, 
where we will derive the phonon quantum estimators. To set the formalism,
Sec.~\ref{pimd} gives a brief overview of the PIMD with Langevin thermostat, i.e. the Path Integral Langevin Dynamics (PILD), whose details can be found in the pioneering work by Ceriotti and coworkers \cite{Ceriotti_2010,Ceriotti_2011}. However, in our work the PILD equation of motions (EOMs) are integrated using the PIOUD algorithm, introduced recently in Ref.~\onlinecite{Mouhat_2017} by some of us. PIOUD is a general
scheme, capable of propagating EOMs with both deterministic forces - such as in DFT - and stochastic forces - such as in the Quantum Monte Carlo (QMC) case -, keeping the temperature constant. PIOUD is very efficient, because it is built upon an optimal number of Trotter factorizations of the Liouvillian operator.\cite{Mouhat_2017}
In Sec.~\ref{pimd_corr_func} we introduce the general formalism to compute quantum mechanical expectation values in PIMD. The remaining Subsections are devoted to phonon evaluation. In Sec.~\ref{classical_estimators} we recall the formula for the phonon computation obtained from a localization criterion for the \emph{classical} velocity-velocity correlation functions\cite{Martinez_2006}. Then, we develop the PIMD extension of the classical correlators by using the Kubo formalism. In Sec.~\ref{force_force}, the quantum force-force phonon estimators are derived, while in Sec.~\ref{disp_disp} we provide the derivation of the phonons calculation based on quantum displacement-displacement correlators.
For the sake of clarity, all derivations are first done considering an isolated system, i.e. a molecule, thus neglecting the periodicity 
of a crystalline system. The generalization to periodic systems is carried out in Sec.~\ref{crystal}. Finally, in Sec.~\ref{sec:ff_dxdx_pimd} we discuss the advantages 
of computing phonons by complying with the localization principle scheme, once compared with Kubo formulas that do not fulfill it.

\subsection{Path Integral Molecular Dynamics}
\label{pimd}
We consider a system of $N$ distinguishable particles.
Using a Trotter factorization of the trace, the quantum mechanical partition function $Z=\text{tr}[e^{-\beta H}]$ can be 
written
as \cite{Feynman_1965,Ceriotti_2010}
\begin{equation}\label{eq:z_pimd}
    Z =
    \lim_{P\rightarrow \infty}
    \frac{1}{(2 \pi \hbar)^f} \int d^f \mathbf{p} \int d^f \mathbf{x} \ e^{-\beta_P H_P(\mathbf{x},\mathbf{p})},
\end{equation}
with $P$ the number of beads, $f=N \cdot P$, $\beta_P=\beta/P$, $1/\beta=k_b T$, and where $\mathbf{x}$ and $\mathbf{p}$ are the $3NP$-dimensional vectors of positions and momenta respectively. In practice, Eq.~(\ref{eq:z_pimd}) is evaluated 
using a discretization approximation with $P$ finite. However, $P$ is taken large enough to guarantee the convergence of the integral.
The PIMD Hamiltonian $H_P (\mathbf{x},\mathbf{p})$ in Eq.~(\ref{eq:z_pimd}) is 
\begin{eqnarray}\label{eq:h_pimd}
    H_P(\mathbf{x},\mathbf{p})=\sum_{i=1}^{3N} \sum_{j=1}^{P} && \left( \frac{[p_i^{(j)}]^2}{2 m_i} + \frac{1}{2}m_i \omega_P^2 [x_i^{(j)}-x_i^{(j-1)}]^2 \right) \nonumber\\ && + \sum_{j=1}^{P} V(x_1^{(j)},...,x_{3N}^{(j)}),
\end{eqnarray}
where $\omega_P=1/(\beta_P \hbar)$ and $x_i^0 = x_i^P$. 
Henceforth, we set $\hbar=1$. The Hamiltonian $H_P$ is defined in an extended phase-space that consists of $P$ images (beads) of the physical N-particle system connected each other through harmonic potentials, while the particles within every image interact through the potential $V$. This maps the quantum system onto a classical model of interacting ring polymers \cite{Feynman_1965,Ceperley_1995}. 

For later purposes, it is convenient to split the extended Hamiltonian as $H_P(\mathbf{x},\mathbf{p})=T_{P}(\mathbf{p})+V_{P}(\mathbf{x})$, where 
\begin{eqnarray}\label{eq:h_pimd_decomp}
&&V_{P}=\frac{1}{2}\sum_{i=1}^{3N} \sum_{j=1}^{P}   m_i \omega_P^2 [x_i^{(j)}-x_i^{(j-1)}]^2 + \sum_{j=1}^{P} V(x_1^{(j)},...,x_{3N}^{(j)}),
 \nonumber\\
&&T_{P}=\sum_{i=1}^{3N} \sum_{j=1}^{P}   \frac{[p_i^{(j)}]^2}{2 m_i}.
\end{eqnarray}
The choice of $m_i$ equal to the physical masses of the particles \cite{Parrinello_1984} and the
EOMs
generated by Eq.~(\ref{eq:h_pimd}) in the extended phase-space correspond to the RPMD approximation. Therein, we neglect the statistics of quantum particles, which are treated as Boltzmannons. Morever, the RPMD trajectories are only meant to sample the stationary quantum thermal distribution, and do not represent a \emph{real time} dynamics \cite{Ceriotti_2010}.

In order to keep the temperature constant during the simulations, we used 
a Langevin thermostat and 
the PIOUD algorithm \cite{Mouhat_2017}. It is worth noting that different 
algorithms 
have been developed to integrate the nuclear EOMs with built-in Langevin thermostats, 
such as the path integral Langevin equation \cite{Ceriotti_2010} (PILE) or the generalized Langevin equation \cite{Ceriotti_2010, Ceriotti_2011, Rossi_2017} (GLE).
At variance with PILE, PIOUD 
integrates exactly the thermal Brownian motion of a quantum particle. In particular,  
while the PILE Liouvillian operator is split 
into normal modes evolution of harmonic oscillators and Langevin thermostatting,
in PIOUD there is a single Liouvillian propagation step that includes both 
processes.
It corresponds to an Ornstein Uhlenbeck dynamics for the 
 $\mathbf{x}$ and $\mathbf{p}$ coordinates, which saves a Trotter breakup.
This nice 
PIOUD feature
allows one to use larger integration time steps without reducing the number of beads, that is of paramount importance for saving CPU time.
On the other hand, GLE is well designed for faster convergence with the number of beads, but it does not fulfill the fluctuation-dissipation theorem.
Therefore, the PIOUD integration scheme is 
the most suitable
for our purposes.

The Born-Oppenheimer potential energy surface and nuclear forces
are computed at each time step by using the 
DFT-based Quantum Espresso \cite{qe1} 
engine for solving the electronic 
Schrödinger equation\cite{Hohenberg_1964,Kohn_1965}. Thus, our approach is fully \emph{ab initio}.

\subsection{Quantum mechanical observables and PIMD correlation functions}
\label{pimd_corr_func}
In the PIMD framework, the quantum mechanical expectation value $\langle A \rangle = \frac{1}{Z} \text{tr}[e^{-\beta H} \hat{A}]$ takes the expression \cite{Craig_2004,Hele_2017} 
\begin{equation}\label{eq:pimd_O_exp_value}
    \langle A \rangle = \int d^f \mathbf{p} \int d^f \mathbf{x} \ \frac{e^{-\beta_P H_P(\mathbf{x},\mathbf{p})}}{Z} \mathsf{A}(\mathbf{x}) \equiv 
    \llrrangle{\mathsf{A}},
\end{equation}
where $\llrrangle{\cdots}$ indicates the average over the paths that sample the statistical distribution of 
$H_P (\mathbf{x},\mathbf{p})$, and
\begin{equation}
\label{eq:bead_average}
   \mathsf{A}(\mathbf{x})=\frac{1}{P}\sum^P_{j=1} A(\mathbf{x}^{(j)}),
\end{equation}
 is the bead-averaged operator, $\mathbf{x}^{(j)}$ being the 3N-dimensional vector of coordinates of the $j$-th image. In Eq.~(\ref{eq:pimd_O_exp_value}) we have assumed that the operator $\hat{A}$ is purely position-dependent, and we have used the cyclic invariance of the trace.
The trajectories generated by the PIMD EOMs, $\{\mathbf{x}^{(j)}(t)\}_{j=1,\ldots,P}$,
are used to 
evaluate the integrals $\llrrangle{\cdots}$, such as the one above.

The time-correlation between two observables $A$ and $B$
is defined 
as:
\begin{equation}\label{eq:true_quantum}
\small
    c_{AB}(t)=\frac{1}{Z}\text{tr} \Big[ e^{-\beta \hat{H}} \hat{A} \ e^{i \hat{H}t} \hat{B} \  e^{-i \hat{H}t}  \Big],
\end{equation}
while the Kubo-transformed version of Eq.~(\ref{eq:true_quantum}) reads:
\cite{Kubo_1957,Craig_2004,Witt_2009,Hele_2016,Hele_2017}
\begin{eqnarray}\label{eq:kubo_time}
    \tilde{c}_{AB}(t)&&=\frac{1}{\beta Z}\int_{0}^{\beta} d \lambda \ \text{tr} \Big[ e^{-(\beta -\lambda)\hat{H}} \hat{A} \ e^{-\lambda \hat{H}} e^{i \hat{H}t} \hat{B} \  e^{-i \hat{H}t}  \Big]. 
\end{eqnarray}
Their Fourier transforms are related each other via the Equation:
\begin{equation}\label{eq:stand_kubo_fourier}
\small
    C_{AB}(\omega)=\frac{\beta \omega}{1-e^{-\beta \omega}} \tilde{C}_{AB}(\omega),
\end{equation}
where $C_{AB}(\omega)$ ($\tilde{C}_{AB}(\omega)$) is the Fourier transform of $c_{AB}(t)$ ($\tilde{c}_{AB}(t)$).
The 
relation in Eq.~(\ref{eq:stand_kubo_fourier}) shows that 
the time dependent Kubo-transformed correlation function of Eq.~(\ref{eq:kubo_time}) carries the same information as the one in Eq.~(\ref{eq:true_quantum}). Nevertheless, one can show \cite{Craig_2004} that $\tilde{c}_{AB}$ is real and even with respect to $t$, a property shared with classical MD correlation functions. This is
in contrast to $c_{AB}$ which, in general, is a complex function. 
The Kubo-transformed correlation function can be derived using linear response theory, \cite{Yamamoto_1960,Hele_2017} and its time dependence is accessible over a short time-scale through PIMD simulations, due to its parallelism with classical correlation functions.
In this work, however, we will just focus on instantaneous correlations. Hereafter, we will refer only to $t=0$ correlation functions and, thus, for the sake of clarity we will drop the time-dependence in our notation, i.e. $\tilde{c}_{AB} \equiv \tilde{c}_{AB}(0)$.
Within PIMD, the zero-time $A$-$B$ correlation function and its Kubo transform read\cite{Craig_2004,Habershon_2013,Hele_2016}
\begin{eqnarray}
\label{eq:pimd_AB}    
   c_{AB} & = & 
    \int d^f \mathbf{p} \int d^f \mathbf{x} \ \frac{e^{-\beta_P H_P(\mathbf{x},\mathbf{p})}}{Z} 
    \frac{1}{P} \sum_{j=1}^P A(\mathbf{x}^{(j)}) B(\mathbf{x}^{(j)}), 
    \\
    \label{eq:pimd_AB_exp_value}
    \tilde{c}_{AB} & = & \int d^f \mathbf{p} \int d^f \mathbf{x} \ \frac{e^{-\beta_P H_P(\mathbf{x},\mathbf{p})}}{Z}
    \mathsf{A}(\mathbf{x})  \mathsf{B}(\mathbf{x}) ,
\end{eqnarray}
respectively. Note that while $c_{AB}$ is the average over products of equal-image operators, the Kubo-transformed correlator $\tilde{c}_{AB}$ includes all product terms, both diagonal and off-diagonal in the bead index. Analogously to Eq.~(\ref{eq:pimd_O_exp_value}), a shorthand notation for Eq.~(\ref{eq:pimd_AB_exp_value}) is $\tilde{c}_{AB}\equiv\llrrangle{ \mathsf{A} \mathsf{B}}$.
The latter will be the building block of our quantum estimators.

\subsection{Classical Phonon Estimators}
\label{classical_estimators}
We now focus on observables related to the evaluation of phonon frequencies. We start from the N-particle classical Hamiltonian, for which anharmonic vibrational theories have 
already been extensively developed in previous works \cite{Martinez_2006,Hellman_2011,Hellman_2013,Paulatto_2013}.
This formally corresponds to the $P=1$ limiting case of Eq.~(\ref{eq:h_pimd}), although the classical situation is substantially different from the quantum case, as Eq.~(\ref{eq:h_pimd}) becomes the \emph{physical}, albeit classical, Hamiltonian, and its EOMs represent a \emph{real time} dynamics.
In this limit, $T_{P=1}=K$, i.e. the classical kinetic energy, and $V_{P=1}=V$, i.e. the inter-particle potential.

The key quantity to compute phonons and 
other derived quantities, such as the vibrational entropy and vibrational free energies, is the force constant matrix, defined as the second derivative of the potential energy $V$ with respect to the displacement of atoms from their equilibrium configurations. Using Cartesian coordinates (as we will do in the rest of the paper), 
the force constant matrix at zero temperature is $\bar{V}_{i_1 i_2} = \frac{\partial^2 V}{\partial x_{i_1} \partial x_{i_2}}\Big \vert_{\mathbf{\bar{x}}}$, where $\mathbf{\bar{x}}$ is the potential energy minimum and $i_1,i_2=1,...,3N$ are the collective Cartesian indices.
More generally, if the temperature is different from zero, the force constant matrix can be defined as
\begin{equation}
\label{eq:force_constant_matrix}
\bar{V}_{i_1 i_2} \equiv \Big\langle \frac{\partial^2 V}{\partial x_{i_1} \partial x_{i_2}} \Big\rangle, 
\end{equation}
where the brackets indicate the average over a statistical ensemble. The matrix $\bar{V}_{i_1 i_2}$ enters the standard eigenvalue problem to compute phonon frequencies:
\begin{equation}\label{eq:phonon_clas}
\bar{V}_{i_1 i_2} Y_{i_2 i_3}=
\omega^2_{i_3} \  m_{i_1} \ Y_{i_1 i_3},
\end{equation}
where $\omega_k$ is the frequency of the $k$-th normal mode, and $Y_{i_2 i_3}$ is the matrix containing the phonon pattern for each normal mode. 

The usual scheme to get $\bar{V}$ at zero temperature is to 
compute the second derivatives numerically around the equilibrium geometry. 
This is the main idea behind the most popular approaches,
such as 
the frozen phonon approximation\cite{Parlinski_1997,Togo_2015} - where the derivatives are computed explicitly -, 
and density functional perturbation theory (DFPT), \cite{Gonze_1997,Baroni_2001} which evaluates second derivatives through response functions.

In this work we will deal with canonical ensemble simulations.
To extract the force constant matrix $\bar{V}$ in this situation, one 
exploits 
an exact relation between force fluctuations and $\bar{V}$, which holds at statistical equilibrium, namely \cite{Martinez_2006,Pereverzev_2015}
\begin{eqnarray}\label{eq:d2vdx2_class_ff}
\langle F_{i_1} F_{i_2} \rangle
&& = \int d^N \mathbf{p} \ e^{- \beta K} \int d^N \mathbf{x} \frac{\partial V}{\partial x_{i_1}} \frac{\partial V}{\partial x_{i_2}} \frac{e^{- \beta V}}{Z} \nonumber\\
&& = \frac{1}{\beta}
\bar{V}_{i_1 i_2} 
\end{eqnarray}
where $Z$ is
the classical partition function and $F_i$ is the force acting on the $i$-th degree of freedom. The above relation, derived integrating by parts with respect to $x_{i_2}$,
is a recipe to obtain the force constant matrix at finite temperature through the evaluation of force-force correlators (i.e. the force covariance matrix). Therefore, it does not require the explicit minimization of the energy. Furthermore, it is worth noting
that Eq.~(\ref{eq:d2vdx2_class_ff}) is very general, because it is fulfilled for any physical potential $V$.

In the context of classical MD simulations, a more general
expression 
than the one of Eq.~(\ref{eq:phonon_clas}) has been derived in Refs.~\onlinecite{Martinez_2006,Gaigeot_2007}, by 
using a 
localization criterion for the Fourier-transformed velocity-velocity
correlation functions. 
This localization principle requires that the power spectrum of the 
position coordinates
is maximally localized in frequency for the effective normal modes. It generalizes the normal mode analysis to anharmonic systems, where the phonon modes acquire a sizeable broadening. 
Following the localization principle, and by making use of the stationarity of correlation functions, 
it is possible to derive a generalized eigenvalue problem where the auto-correlations 
are taken in the zero-time limit. It reads\cite{Martinez_2006}
\begin{equation}\label{eq:ff_class}
    \langle F_{i_1} F_{i_2} \rangle Y_{i_2 i_3}= \omega^2_{i_3} \langle p_{i_1} p_{i_2} \rangle Y_{i_2 i_3},
\end{equation}
where $ \omega^2_{i_3}$ are the squared phonons frequencies. By means of Eq.~(\ref{eq:d2vdx2_class_ff}), 
Eq.~(\ref{eq:ff_class})
looks very similar to
Eq.~(\ref{eq:phonon_clas}), apart from the momentum correlator on its right-hand side. However, one can derive Eq.~(\ref{eq:phonon_clas}) from the most general Eq.~(\ref{eq:ff_class}) thanks to the relation:
\begin{equation}\label{eq:pipj_clas}
  \langle p_{i_1} p_{i_2} \rangle=k_b T m_{i_1} \delta_{i_1 i_2},  
\end{equation}
which holds at thermodynamic equilibrium, when the momenta of the $3N$ degrees of freedom are decorrelated and fulfill the equipartition theorem.

Another way to 
obtain the force constant matrix
for a system at thermal equilibrium is based on
the displacement-displacement correlation function: 
\begin{eqnarray}\label{eq:dxi_dxj}
\langle  \delta x_{i_1} \delta x_{i_2} \rangle &&\approx \int d^N \mathbf{p} \ e^{- \beta K} \int d^N \mathbf{x} \ \delta x_{i_1} \delta x_{i_2} \frac{e^{- \frac{\beta}{2}\delta\mathbf{x}^T \cdot \bar{V} \cdot \delta\mathbf{x}}}{Z} \nonumber\\
&&\approx \beta \left[ \bar{V}^{-1} \right]_{i_1 i_2},
\end{eqnarray}
where $\delta \mathbf{x}=\mathbf{x}-\mathbf{\bar{x}}$ is the nuclear displacement from the equilibrium geometry.
At variance with Eq.~(\ref{eq:d2vdx2_class_ff}), in Eq.~(\ref{eq:dxi_dxj}) 
the identity is only approximated, because the relation with $\bar{V}$ relies on the truncation of the Taylor expansion of 
$V$ at the second order in $\delta x$, namely $V=V_0+\frac{1}{2}\frac{\partial^2 V}{\partial x_{i_1} \partial x_{i_2}}\Big\vert_{\mathbf{\bar{x}}}\delta x_{i_1} \delta x_{i_2} + o (\delta x^3)$. The identity in Eq.~(\ref{eq:dxi_dxj}) becomes exact in the small temperature limit, or if the potential $V$ is harmonic.
Nonetheless, using Eq.~(\ref{eq:dxi_dxj}) to estimate the force constant matrix and inserting it into Eq.~(\ref{eq:phonon_clas}) corresponds to 
the 
Principal Mode Analysis method \cite{Brooks_1995,Wheeler_2003}. 
Analogously to
the force correlation functions, the localization principle of Ref.~\onlinecite{Martinez_2006} leads to another generalized eigenvalue version of Eq.~(\ref{eq:phonon_clas}) that can be obtained from the zero-time displacement-displacement correlation function:
\begin{equation}\label{eq:dxdx_class}
    \left[ \langle \delta\mathbf{x} \delta\mathbf{x}^T \rangle^{-1} \right]_{i_1 i_2 } W_{i_2 i_3}= 
    \omega^2_{i_3}
    \left[ \langle \dot{\mathbf{x}} \dot{\mathbf{x}}^T \rangle^{-1} \right]_{i_1 i_2} W_{i_2 i_3},
\end{equation}
where $W$ is the matrix containing the patterns of the principal modes. 
By exploiting 
the equipartition theorem for velocities:
\begin{equation}
    \left[ \langle \dot{\mathbf{x}} \dot{\mathbf{x}}^T \rangle^{-1} \right]_{i_1 i_2}=
    m_{i_1} \beta \delta_{i_1 i_2},
\end{equation}
one can derive the standard eigenvalue Eq.~(\ref{eq:phonon_clas}) from the generalized eigenvalue Eq.~(\ref{eq:dxdx_class}).
It is important to notice that for harmonic systems, or in the small temperature limit, Eqs. (\ref{eq:ff_class}) and (\ref{eq:dxdx_class}) correspond to the same generalized eigenvalue problem and, thus, yield the same frequencies at thermodynamic equilibrium. At zero temperature both principal mode and normal mode analysis yield the same eigenvalues, which in this limit correspond to the harmonic frequencies.

\subsection{Quantum force-force estimator}
\label{force_force}
To quantize the classical approach we reviewed in Sec.~\ref{classical_estimators}, let us take into account the force constant matrix defined in Eq.~(\ref{eq:force_constant_matrix}). According to its definition, $\bar{V}_{i_1 i_2}$ can be seen as an instantaneous two-particle correlation function that weights the coordinates of particles $i_1$ and $i_2$ by the amplitude $\frac{\partial^2 V}{\partial x_{i_1} \partial x_{i_2}}$. Therefore, we can extend its definition within the PIMD framework either using Eq.~(\ref{eq:pimd_AB}), or its Kubo transform in Eq.~(\ref{eq:pimd_AB_exp_value}). Since we will show that the most appropriate quantum estimators to compute phonons in PIMD simulations are those based on Kubo-transformed correlation functions, we define the PIMD force constant matrix as:
\begin{equation}
\label{eq:PIMD_force_constant_matrix}
\bar{V}_{i_1 i_2} \equiv
\llrrangle{\frac{1}{P^2}  \sum_{j_1,j_2=1}^{P} \frac{\partial^2 \mathsf{V}}{\partial x^{(j_1)}_{i_1} \partial x^{(j_2)}_{i_2}}}= \llrrangle{\bar{\mathsf{V}}_{i_1 i_2}},
\end{equation}
where the rightmost expression is a shorthand notation, with $\bar{\mathsf{V}}_{ij}$ the Kubo-averaged Hessian matrix of $\mathsf{V}(\mathbf{x})=\frac{1}{P}\sum_{j=1}^P V(\mathbf{x}^{(j)})$.

By also defining the quantum force fluctuation by means of a Kubo-transformed correlation function, namely $\tilde{c}_{F_{i_1}F_{i_2}}=\llrrangle{\mathsf{F}_{i_1} \mathsf{F}_{i_2}}$, 
we prove that the following relation holds (see App.~\ref{app:dxdx} for the full derivation):
\begin{equation}\label{eq:average_ff_pimd} 
 \llrrangle{\mathsf{F}_{i_1} \mathsf{F}_{i_2}} = \frac{1}{\beta} \llrrangle{\bar{\mathsf{V}}_{i_1 i_2}}.
\end{equation}
This is the exact PIMD analogue of 
Eq.~(\ref{eq:d2vdx2_class_ff}), and - like in the classical case - the above relation 
is general, because it does not depend on the form of the potential $V$. If a similar derivation were carried out by using PIMD correlation functions as per Eq.~(\ref{eq:pimd_AB}), Eq.~(\ref{eq:average_ff_pimd}) would be broken, as shown in App.~\ref{app:dxdx} as well, and no quantum analogue of the classical relation would be found. This strongly supports the idea that the Kubo-transformed correlation functions are the most suitable extensions of the corresponding classical quantities.

Furthermore, in a quantum system, the zero-time Kubo-transformed autocorrelation for Cartesian momenta fulfills the equipartition theorem, i.e. $\tilde{c}_{p_{i_1} p_{i_2}}= m_{i_1} k_b T \delta_{i_1 i_2}$. This relation holds also for the momenta of the effective classical system of ring polymers. Thus, given the classical localization principle of the effective normal modes \cite{Martinez_2006} in terms of force autocorrelation function, these quantum relations enable us to write down the corresponding PIMD generalized eigenvalue problem for phonons evaluation:
\begin{equation}\label{eq:ff_pimd}
    \llrrangle{\mathsf{F}_{i_1} \mathsf{F}_{i_2}} Y_{i_2,i_3}= \omega_{FF,i_3}^2 \llrrangle{ \mathsf{p}_{i_1} \mathsf{p}_{i_2}} Y_{i_2,i_3},
\end{equation}
where both $\mathsf{F}$ and $\mathsf{p}$ observables are averaged over the whole ring polymers, as defined in Eq.~(\ref{eq:bead_average}). Like in the classical case, Eq.~(\ref{eq:ff_pimd}) is more general than the standard eigenvalue problem, which can  be readily derived from the definition of the PIMD force constant matrix in Eq.~(\ref{eq:PIMD_force_constant_matrix}), and reads:
\begin{equation}\label{eq:ff_pimd_direct_dyn}
    \llrrangle{ 
    \bar{\mathsf{V}}_{i_1 i_2}} Y_{i_2,i_3}= \omega^2_{i_3} \ m_{i_1} \ Y_{i_1,i_3}.
\end{equation}
The generalized eigenvalue Eq.~(\ref{eq:ff_pimd}) reduces to the standard eigenvalue Eq.~(\ref{eq:ff_pimd_direct_dyn}) in the 
thermodynamic equilibrium limit
of the extended phase-space with $3NP$ degrees of freedom. Therefore, the $\omega_{FF,i}$ eigenvalue is the fundamental frequency of the $i$-th normal mode. 
However, Eq.~(\ref{eq:ff_pimd}) does not require 
perfect equipartition, in contrast to Eq.~(\ref{eq:ff_pimd_direct_dyn}). This will turn out to be a great advantage for a much faster convergence and a more precise evaluation of the phonon modes, when the generalized eigenvalue problem is applied to the quantum case.


In Sec.~\ref{sec:ff_dxdx_pimd}, by using 1D model potentials, we will examine 
the equilibration time and the time step bias affecting the phonon frequencies convergence during the PIMD EOMs evolution.
By comparing the results obtained with both eigenvalue problems Eqs.~(\ref{eq:ff_pimd}) and (\ref{eq:ff_pimd_direct_dyn}), we will demonstrate the better performance of Eq.~(\ref{eq:ff_pimd}), which 
improves the phonon estimation efficiency by at least one order of magnitude with respect to the standard approach of Eq.~(\ref{eq:ff_pimd_direct_dyn}).

\subsection{Quantum displacement-displacement estimator}
\label{disp_disp}


As done for the force-force correlation functions in Sec.~\ref{force_force}, we want here to extend the classical displacement-displacement correlators in order to compute phonons in a quantum system within the PIMD framework. As already seen for the force-force estimators, the most appropriate quantum correlation functions, which preserve the relationships found in the classical situation, are the ones based on the Kubo transform. Thus, we proceed by quantizing 
the classical autocorrelations present in the principal mode analysis and in the classical generalized eigenvalue problem of Eq.~(\ref{eq:dxdx_class}).

In the PIMD notation, the displacement-displacement and velocity-velocity autocorrelation matrix elements are given by
\begin{eqnarray}
    && \llrrangle{ \delta \mathsf{x}_{i_1} \delta \mathsf{x}_{i_2}} = \llrrangle{\left( \frac{1}{P}\sum_{j_1=1}^P \delta x^{(j_1)}_{i_1} \right)  \left( \frac{1}{P}\sum_{j_2=1}^P \delta x^{(j_2)}_{i_2} \right)}, \nonumber \\
    && \llrrangle{\mathsf{\dot{x}}_{i_1} \mathsf{\dot{x}}_{i_2}} = \llrrangle{\left( \frac{1}{P} \sum_{j_1=1}^P \dot{x}^{(j_1)}_{i_1} \right)  \left( \frac{1}{P} \sum_{j_2=1}^P \dot{x}^{(j_2)}_{i_2} \right)}.
\end{eqnarray}
Thus, following the same path as the force-force case, by extending Eq.~(\ref{eq:dxdx_class}) into the PIMD case, one 
obtains
\begin{equation}\label{eq:dxdx_pimd}
    \left[ 
    \llrrangle{\delta \bm{\mathsf{x}} \delta \bm{\mathsf{x}}^T}^{-1} \right]_{i_1,i_2} W_{i_2,i_3}=\omega_{\delta x \delta x,i_3}^2 \left[\llrrangle{ \bm{\mathsf{\dot{x}}} \bm{\mathsf{\dot{x}}}^T}^{-1} \right]_{i_1,i_2} W_{i_2,i_3},
\end{equation}
where we adopted the same notation as the classical case, 
but the classical correlation functions are replaced by the Kubo-transformed quantum auto-correlations.
Like in the classical case, it is true that $\tilde{c}_{\dot{x}_i \dot{x}_j}=\delta_{i,j} k_b T/m_i$. However, at variance with the force-force estimator of Eq.~(\ref{eq:average_ff_pimd}), the Kubo displacement-displacement correlator is not exactly related to the Kubo force constant matrix. This is very similar to the classical behavior, where only the classical force auto-correlation can be exactly expressed in terms of $\bar{V}_{i_1 i_2}$ (Eq.~(\ref{eq:d2vdx2_class_ff})), while the relation between the classical displacement-displacement correlator and $\bar{V}_{i_1 i_2}$ is approximated (Eq.~(\ref{eq:dxi_dxj})). However, at variance with the classical framework, the broken relation between the Kubo-transformed displacement correlator and the Kubo-transformed force constant matrix hides a deeper physical meaning for the eigenvalues $\omega_{\delta x \delta x,i}$ in Eq.~(\ref{eq:dxdx_pimd}). Indeed, such eigenvalues are no longer an estimation of the fundamental frequencies of the system, but they are more directly related to the first energy excitation of the phonon modes.

To show this property in a simple and transparent situation, let us
focus on a 1D system with Hamiltonian $\hat{H}=\frac{\hat{p}^2}{2m}+\hat{V}(x)$. To be more explicit, here we will use the notation of the true quantum Kubo-transformed correlation functions instead of their PIMD versions. With our aim,
it is useful to decompose Eq.~(\ref{eq:kubo_time}) at zero time in terms of the complete set of quantum eigenstates $\{|n\rangle\}$ of the time-independent Hamiltonian $\hat{H}$:
\begin{eqnarray}\label{eq:kubo_decomp}
    \tilde{c}_{AB}&&=\frac{1}{\beta Z}\int_{0}^{\beta} d \lambda \ \text{tr} \Big[ e^{-(\beta -\lambda)\hat{H}} \hat{A} \ e^{-\lambda \hat{H}} \hat{B} \ \Big] \nonumber\\
    && =\frac{1}{\beta Z} \sum_{n,m} e^{- \beta E_n} A_{nm} B_{mn} \frac{1-e^{-\beta \omega_{m,n}}}{\omega_{m,n}} \nonumber \\
    && = \frac{1}{\beta Z} \sum_{n,m} \tilde{c}^{(nm)}_{AB},
\end{eqnarray}
where $\omega_{m,n}=E_m - E_n$ is the energy difference between 
the $m$-th and $n$-th quantum levels, and $A_{mn}=\langle m \vert \hat{A} \vert n \rangle$.
In the 1D case, Eq.~(\ref{eq:dxdx_pimd}) 
can be rewritten as:
\begin{equation}\label{eq:cc_expl_1d}
\small
     \omega^2_{\delta x \delta x}=\frac{\tilde{c}_{\dot{x} \dot{x}}} {\tilde{c}_{\delta x \delta x}}=\frac{\sum_{l,n} \omega^2_{l,n} \cdot \tilde{c}^{(l,n)}_{\delta x \delta x}}{\sum_{l,n} \tilde{c}^{(l,n)}_{\delta x \delta x}}=\sum_{l,n} \omega^2_{l,n} \cdot \tilde{d}^{(l,n)}_{\delta x \delta x},
\end{equation}
where we employed the 
correlation functions 
decomposition 
in terms 
of the Hamiltonian eigenstates,
as detailed in Eq.~(\ref{eq:kubo_decomp}), and we used the quantum mechanical equivalence $\langle l | \hat{\dot{x}}_{i} | n \rangle = i \cdot  \langle l | \hat{x}_{i} | n \rangle \cdot \omega_{l,n}$. At the rightmost-hand side of Eq.~(\ref{eq:dxdx_pimd}), we defined the coefficients
\begin{equation}
    \tilde{d}^{(l,n)}_{\delta x \delta x} = \frac{ \tilde{c}^{(l,n)}_{\delta x \delta x}}{\sum_{l,n} \tilde{c}^{(l,n)}_{\delta x \delta x}},
\end{equation}
which are normalized to 1, once summed over $l$ and $n$ (i.e. $\sum_{l,n} \tilde{d}^{(l,n)}_{\delta x \delta x} = 1$). From Eq.~(\ref{eq:kubo_decomp}), we observe that the quantum displacement-displacement estimator 
is a sum of 
squared-transitions between the eigenstates of the system. In particular the coefficients $\tilde{d}^{(l,n)}_{\delta x \delta x}$ are proportional to $\propto \frac{ | \langle l | \left( \hat{x} -\bar{x} \right) | n \rangle |^2}{\omega_{n,l}}$. Considering the $T \to 0$ limit, where only the ground state is populated, the coefficients become (see appendix \ref{app:details_c}):
\begin{equation}\label{eq:1d_expl}
    \tilde{d}^{(l,n)}_{\delta x \delta x} \simeq \tilde{d}^{(0,n)}_{\delta x \delta x} \propto \frac{ | \langle 0 | \left( \hat{x} -\bar{x} \right) | n \rangle |^2}{\omega_{n,0}}, 
\end{equation}
where $n = \{1,..,\infty\}$ runs over the excited-states indices, and $|0 \rangle$ represents the ground state.
\begin{figure}[h!]
\includegraphics[width=0.5\textwidth]{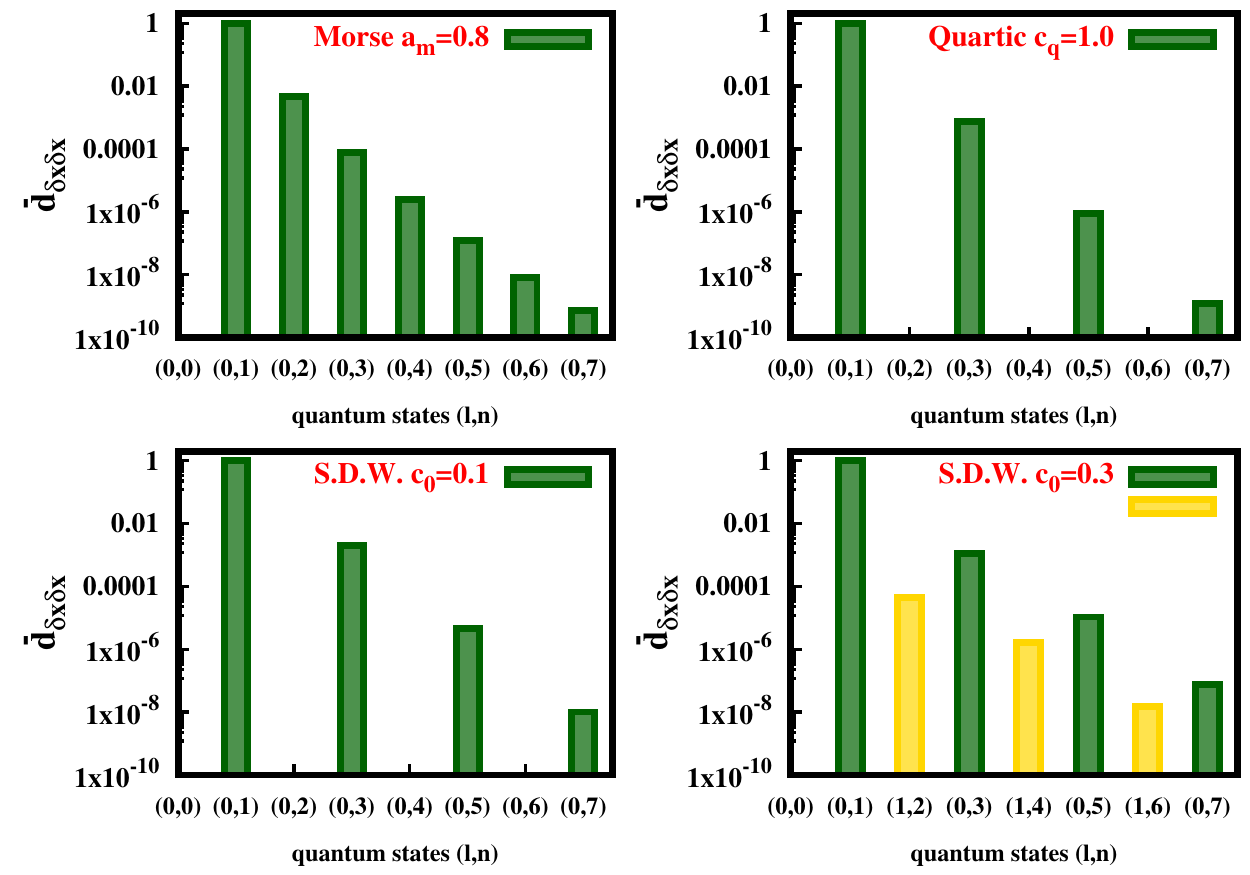}
\caption{\label{fig:weights1d} Semi-log plot of the quantum displacement-displacement estimator weights $\tilde{d}^{(0,n)}_{\delta x \delta x}$ at 20 K for the Morse potential (top-left panel), quartic (top-right panel) and symmetric double well with different parameters (bottom-left and bottom-right panels) with respect to quantum state transitions. The $n \rightarrow m$ transition is labelled as $(n,m)$ on the $x$ axis. At low temperature, only transitions from the ground state are significantly different from zero (Eq.~(\ref{eq:1d_expl})). In the case of symmetric double well (S.D.W.) with $c_0=0.3$, the yellow bars correspond to transition from the first excited state, which for high potential barriers is very close to the ground state. A detailed description of the 1D models can be found in Sec.~\ref{sec:results_1d}.}
\end{figure}
By inspecting Eq.~(\ref{eq:1d_expl}), it is straightforward to see that $\tilde{d}^{(0,n)}_{\delta x \delta x}$ is positive definite $\forall n$. Therefore, it can be interpreted as a weighting function for the squared transition energies in Eq.~(\ref{eq:cc_expl_1d}). Moreover, it is a decreasing function of $n$. Indeed, in Eq.~(\ref{eq:1d_expl}) both numerator and $1/\omega_{n,0}$ are decreasing with $n$.
To illustrate and quantify the $n$-dependence of the weights in Eq.~(\ref{eq:1d_expl}), 
in Fig.~\ref{fig:weights1d} 
we plot their values 
for 1D toy-model potentials that will be presented in more detail in Sec.~\ref{sec:results_1d}.
In all cases studied, more than $99\%$ of the total weight belongs to the first energy transition. Moreover, the decay with 
$n$ is exponentially fast. Despite the specific examples reported here, we found that this behaviour is common to all systems 
we have studied
in Sec.~\ref{sec:results}, which are representative of a large and diverse set of anharmonic situations. 
These are therefore universal features, and we can take the displacement-based generalized eigenvalue problem in Eq.~(\ref{eq:dxdx_pimd}) as an accurate way to access the first energy excitations of the phonon spectrum. 

A similar approach for estimating 
phonon excitation energies has also been proposed in Refs.~\onlinecite{Ramirez_1999,Ramirez_2001}, where low-lying phonons excitations are computed by the standard eigenvalue problem that can be derived by plugging $\tilde{c}_{\dot{x}_i \dot{x}_j}=\delta_{i,j} k_b T/m_i$ into Eq.~(\ref{eq:dxdx_pimd}). Therefore, in that case, the equipartition is imposed \emph{a priori}. Instead, in our case, the rigorous quantization of the classical relations allowed us to derive a PIMD generalized eigenvalue problem analogous to the classical localization principle, which does not require a perfect equipartition. Remarkably, we found that
the generalized eigenvalue problem is much more efficient than the standard one in predicting the vibrational frequencies in PIMD, as reported in Sec.~\ref{sec:ff_dxdx_pimd} for the force-force as well as for the displacement-displacement estimators. 

Finally, we notice that for a harmonic potential, the quantum displacement-displacement estimator coincides with the force-force one - as in the classical case -, and the two approaches give exactly the same eigenvalues. Indeed, in the harmonic situation the fundamental frequency yielded by the force-force estimator equals the first excitation energy $\omega_{1,0}$ provided with unitary weight by the displacement-displacement estimator. 
However, we stress once again that, in the general anharmonic case, the physical information carried by the two eigenvalues is different.

\subsection{PIMD estimators for crystalline systems}
\label{crystal}
While for isolated systems like molecules the correlation functions accumulated over the 
MD or PIMD
trajectories 
are usually enough to construct good estimators\footnote{However, in molecular calculations one has to define a proper frame for appropriate internal coordinates}, for crystals one
should exploit
the space group properties to build symmetric force constant and correlation matrices. Furthermore, to access the phonon dispersion 
in the 
Brillouin zone away from the $\Gamma$-point by means of MD or PIMD simulations,
one makes use of a supercell with periodic boundary conditions, where the 
lattice
vector $\mathbf{a}_i$ is repeated $n_i$ times, with $i=\{1,2,3\}$ and $n_i \geq 1$.
%
The set of integers $(n_1,n_2,n_3)$, i.e. the size of the supercell, should be chosen in such a way that the correlation matrix elements between the central atoms 
and those at the supercell
border is negligible. 

When dealing with periodic boundary conditions, the strategy to compute the phonon dispersion relies on the following procedure:
\begin{itemize}
    \item From MD or PIMD trajectories, we build the supercell interatomic correlation matrices in real space for all quantities of interest, i.e. force and momentum autocorrelations for the force-force estimator, displacement and velocity autocorrelations for the displacement-displacement estimator. In order to avoid global translational drifts, at each time step we apply the following pinning strategy for the matrices that are not gauge invariant with respect to translations.
    The reference frame is pinned over the $i$-th atom of the supercell, so that each atomic position and displacement are computed with respect to this atom. We loop over 
    the equivalent atoms of the supercell, $i=\{1,\ldots,M_\textrm{eq}\}$, by repeating the same procedure for each atom $i$, and averaging the auto-correlation functions over the different pinning centers. In this way, the resulting correlation functions - in particular the displacement correlation matrix - will be fully symmetric and gauge invariant.
    \item For each q-point accessible by the supercell size, namely $\mathbf{q}=\left(\frac{h_1}{n_1}\mathbf{b}_1,\frac{h_2}{n_2}\mathbf{b}_2,\frac{h_3}{n_3}\mathbf{b}_3\right)$, where $\mathbf{b}_i$ are the reciprocal lattice vectors and $h_i=\{0,...,n_i-1\}$, we perform a Fourier transform of the different real-space correlation matrices, and we solve the generalized eigenvalue (GEV) problems (Eqs.~(\ref{eq:ff_class}) and (\ref{eq:dxdx_class}) for MD, and Eqs.~(\ref{eq:ff_pimd}) and (\ref{eq:dxdx_pimd}) for PIMD). In particular, 
    the dual forms of 
    the real-space Eqs.~(\ref{eq:ff_pimd}) and (\ref{eq:dxdx_pimd}) read
          \begin{eqnarray}  
          \label{eq:ff_pimd_crystal}
       && \llrrangle{\mathsf{F}_{i_1} \mathsf{F}_{i_2}}\!\! (\mathbf{q}) ~ Y_{i_2; i_3} (\mathbf{q}) =
       \nonumber\\ && =
       \omega_{FF;i_3}^2 (\mathbf{q})   \llrrangle{\mathsf{p}_{i_1} \mathsf{p}_{i_2}}\!\! (\mathbf{q}) ~ Y_{i_2; i_3} (\mathbf{q}),
    \end{eqnarray}
    and
    \begin{eqnarray}
    \label{eq:dxdx_pimd_crystal}
    &&
        \left[ \llrrangle{ \delta\bm{\mathsf{x}} \delta\bm{\mathsf{x}}^T}^{-1}\!\!\!(\mathbf{q}) \right]_{i_1;i_2} \!\! W_{i_2;i_3} (\mathbf{q}) =
        \nonumber
        \\ &&=
        \omega_{\delta x \delta x; i_3}^2 (\mathbf{q}) \left[ \llrrangle{ \bm{\mathsf{\dot{x}}} \bm{\mathsf{\dot{x}}}^T}^{-1}\!\!\!(\mathbf{q}) \right]_{i_1;i_2} \!\! W_{i_2;i_3} (\mathbf{q}), 
    \end{eqnarray}
    respectively, where $i_1,i_2,i_3=\{1,...,3N\}$ are the indices of the unit-cell degrees of freedom.
    
    \item From the solution of the two GEVs, using the eigenvalues $d_i (\mathbf{q})$ and eigenvectors $\vert d_i (\mathbf{q}) \rangle$ of Eq.~(\ref{eq:ff_pimd_crystal}) or Eq.~(\ref{eq:dxdx_pimd_crystal}), we reconstruct for both estimators the true dynamical matrix:
    \begin{equation}
     \label{eq:matdyn_reconstruction}
        \textrm{D}_m(\mathbf{q}) = \sum_i \vert d_i (\mathbf{q}) \rangle d_i (\mathbf{q}) \langle d_i (\mathbf{q}) \vert,
        \end{equation}
    for each q-point.    
    D$_m$ is otherwise 
    defined as the mass-weighted Fourier transform of the force constant matrix. Eq.~(\ref{eq:matdyn_reconstruction}), obtained from the solution of Eq.~(\ref{eq:ff_pimd_crystal}) or Eq.~(\ref{eq:dxdx_pimd_crystal}), is a generalization of the standard definition.
    \item We impose 
    the space-group symmetry relations 
    in q-space by symmetrizing the dynamical matrix elements for each q-point and by creating all the dynamical matrices belonging to the star of a given q-point \cite{Dresselhaus_2008}. This procedure allows one to average the equivalent dynamical matrices eventually created in the stars of the q-points that are sampled independently during the simulation\footnote{To do that we exploit the Quantum Espresso tool $q2qstar.x$}.
    \item We go back to real space by Fourier transforming the fully symmetrized D$_m$
    and by building the real-space force constant matrix, at this stage being fully symmetric. 
    \item
    As last step, we perform an interpolation of the phonon dispersion, by computing and diagonalizing 
    the final dynamical matrix on a finer q-grid.  
\end{itemize}

\subsection{Generalized eigenvalue vs standard eigenvalue problem}\label{sec:ff_dxdx_pimd}
We 
prove
numerically 
that, for phonon calculations based on force autocorrelations, 
the 
GEV problem in Eq.~(\ref{eq:ff_pimd}) 
is much more efficient
than the standard eigenvalue problem in Eq.~(\ref{eq:ff_pimd_direct_dyn}), where the mass matrix has been replaced
by momentum correlations in the former Equation. In the same way, we compare the displacement-displacement GEV in Eq.~(\ref{eq:dxdx_pimd}) with its standard version as well. 
In order to do that, we performed PIMD simulations of a 1D particle bounded by a Morse potential $V(x)=\frac{k}{2 \cdot a_m^2}\cdot (1-e^{-a_m \cdot x})^2$, with $k=0.4837$ a.u. and $a_m=1.1$ bohr$^{-1}$. The temperature $T$ is 100 K, the number of beads $P=80$, while we varied the integration time-step $\Delta t$ from 0.25 fs to 1 fs. The friction parameter $\gamma_0$ of the Langevin thermostat \cite{Mouhat_2017} is 1.46 $\cdot$ 10$^{-3}$ atomic units. This value is the same as in Ref.~\onlinecite{Mouhat_2017}, where it is found to be optimal for both stochastic and deterministic forces, representing a good compromise between diffusion and thermalization rates in systems with strong NQE. The same value for $\gamma_0$ has been used throughout all calculations presented in this work.

\begin{figure*}
\includegraphics[width=0.85\textwidth]{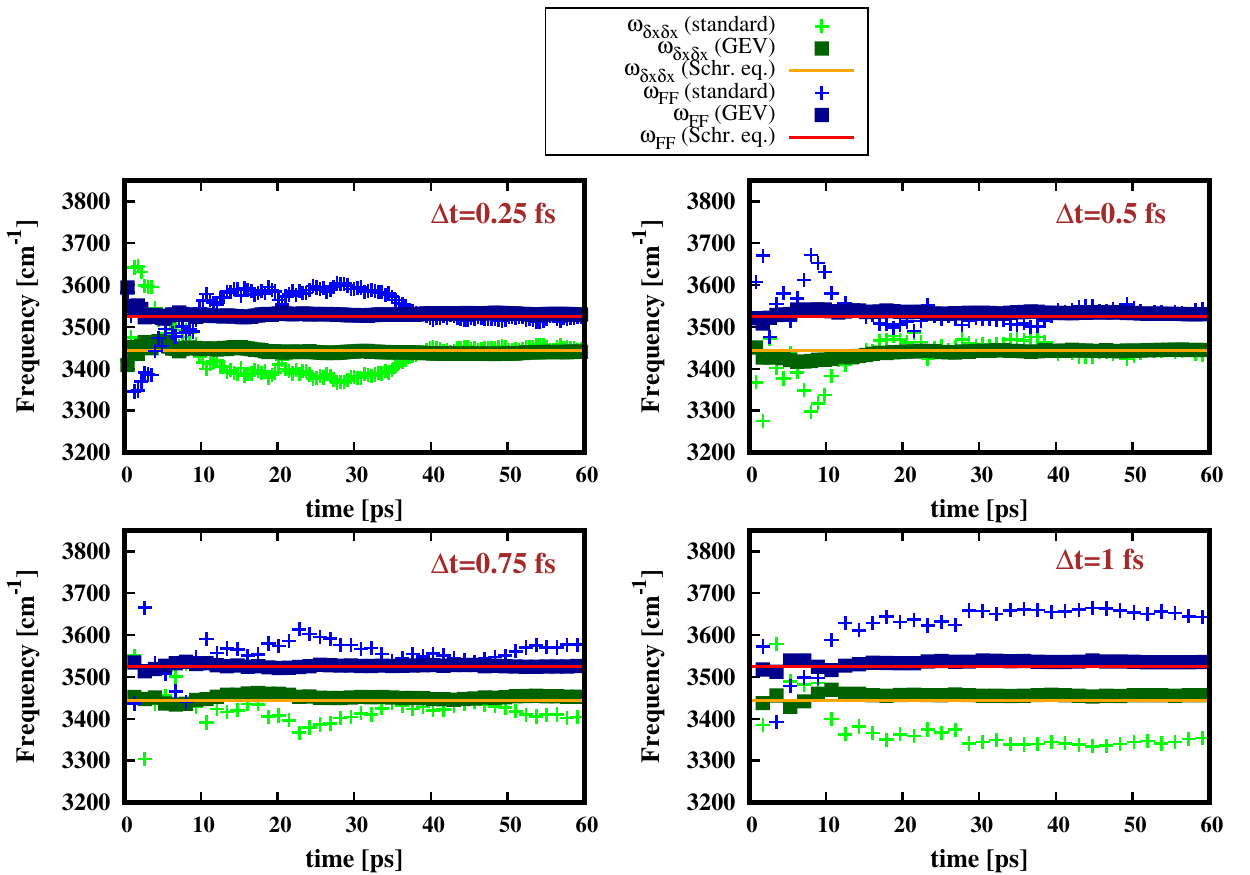}
\caption{\label{fig:ff_dxdx_morse_conv} Convergence of the PIMD phonon eigenvalues for a 1D particle
bounded by a Morse potential at 100 K obtained with different integration time-steps $\Delta t$.
The exact solutions obtained from the numerical Schrödinger equation are given in red for the force-force (fundamental frequency) and orange for the displacement-displacement estimators ($\omega_{1,0}$ excitation energy).}
\end{figure*}

In Fig.~\ref{fig:ff_dxdx_morse_conv} we plot the behavior of the estimated phonon eigenvalues as a function of the PIMD time evolution. We
observe that the frequencies computed from GEV rapidly converge towards the exact values, obtained by diagonalizing numerically the quantum Hamiltonian. 
The convergence is reached after a few ps of dynamics, independently of the $\Delta t$ value.
On the other hand, the solutions of Eq.~(\ref{eq:ff_pimd_direct_dyn}),
dubbed $\omega_{FF} (standard)$, and the one relative to the displacement-displacement estimator,
labelled $\omega_{ \delta x \delta x} (standard)$, show oscillation amplitudes much larger than the ones from the corresponding GEV solutions, in all reported cases. Even worse, for $\Delta t \geq 0.5$ fs, the finite time-step bias become sizeable ($> 50$ cm$^{-1}$), while it is still negligible for the GEV frequencies. 

In App.~\ref{app:gen_vs_sta} we present the same analysis for other two relevant 
cases of anharmonicity: a 1D particle bounded by a quartic potential and the one in a symmetric double well potential. 
The results show the better performance of the GEV estimators with respect to the standard ones in all anharmonic situations. 

We deduce that the improvement given by GEV phonon equations with respect to their standard versions is twofold: firstly they allow to work with a larger time-step $\Delta t$, and secondly their eigenvalues converge on a much shorter time scale.
From the toy models analyzed here, we infer that the gain in $\Delta t$ is at least a factor of 2, while the improvement in the convergence time is even more impressive. We go from about 40 ps in the standard eigenvalue equation to about 4 ps in the GEV. The overall gain in efficiency is therefore larger than one order of magnitude. This is of paramount importance to reduce the global cost of the phonon calculations by PIMD, and to make challenging problems feasible.

Our finding about the performance of the phonon quantum GEV problems whose form has been borrowed from a classical localization principle is particularly important. It is clear that the equipartition theorem implied by the standard eigenvalue problem is hard to reach in the extended $3NP$ space of polymers, where several energy scales coexist. Moreover, it is also clear that nuclear vibrations - particularly the low-energy ones - are very sensitive to the quality of the phase-space sampling. Therefore, the use of GEV for PIMD phonon calculations turns out to be key for an efficient and accurate evaluation of vibrational frequencies. 

\section{Results}
\label{sec:results}
To benchmark the proposed quantum phonon estimators, firstly we studied 1D model potentials, for which we determined the exact vibrational spectrum through the numerical solution of the Schrödinger equation. 

Then, we focused also on a more complex 
2D model potential, where we could tune the interaction of the two degrees of freedom. Also in that case, we computed the exact spectrum, and compared against the PIMD force and displacement autocorrelation approaches for benchmark.

Finally, as real test cases, we studied the phonon dispersions of diamond and of atomic hydrogen at high pressure in the I4$_1$/amd phase. 
\subsection{One-dimensional models}\label{sec:results_1d}
\begin{figure}[h!]
\includegraphics[width=0.5\textwidth]{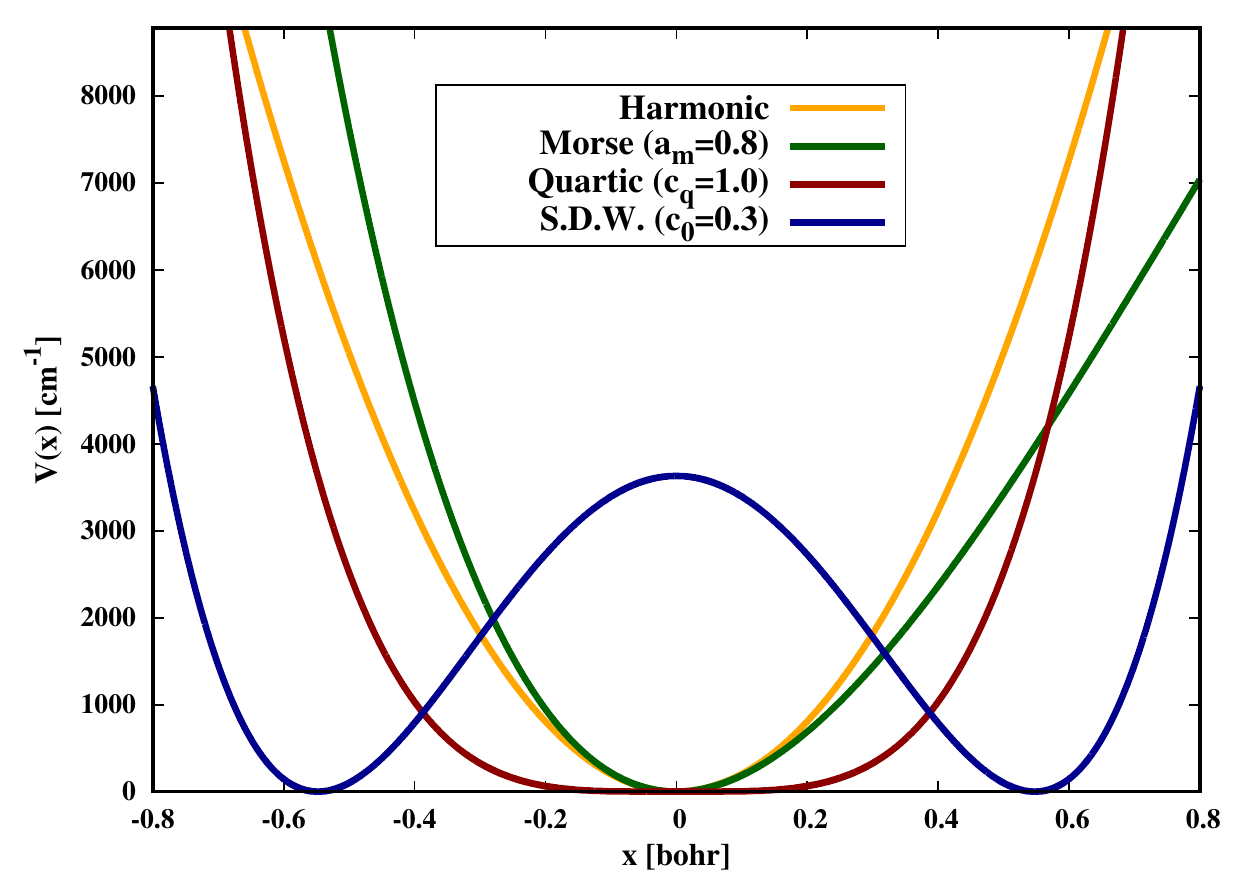}
\caption{\label{fig:pot1d} 1D potentials used to test the behaviour of the quantum estimators. For each potential-type, we plot only the most anharmonic one.}
\end{figure}

For the 1D case, we studied the following potentials, 
listed
in an increasing order of anharmonicity and shown in Fig.~\ref{fig:pot1d}:
\begin{itemize}
\item Harmonic potential: $V(x)= \frac{k}{2}\cdot x^2$;
\item Morse potential: $V(x)= \frac{k}{2 \cdot a_m^2}\cdot (1-e^{-a_m \cdot x})^2$ with $a_m$ = 0.2, 0.4, 0.6, 0.8;
\item Quartic potential: $V(x)=c_q \cdot k \cdot x^4$ with $c_q$ = 0.01, 0.1, 1;
\item Symmetric double well potential: $V(x)=k \cdot (x^2-c_{0})^2$ with $c_{0}$ = 0.05, 0.1, 0.3.
\end{itemize}

Then, as a realistic 1D case, we evaluate the PIMD phonon estimators also for the hydrogen molecule using the accurate potential taken from Ref.~\onlinecite{Kolos_1965}.

In the model potentials, the parameter $k$ is always fixed equal to $0.183736$ atomic units, while the physical mass is chosen equal to the hydrogen atom mass. These choices correspond to an oscillation frequency for the harmonic case of $\sqrt{k/m}=\omega_{harm}=2194.74$ cm$^{-1}$.

The details of Schrödinger equation solutions can be found in App.~\ref{app:1d_2d_sch}.
We have already introduced the notation for the PIMD
eigenvalues obtained from the displacement-displacement ($\omega_{\delta x \delta x}$) and force-force ($\omega_{FF}$) autocorrelations; In the following we call $\frac{\omega_0}{2}$ the exact ground state energy and $\omega_{1,0}$ the exact first excitation energy. 

It is worth noting
that in the harmonic case, all the estimators are equivalent and - as expected - give exactly the value of $\omega_{harm}$ as fundamental frequency independently of the temperature. In such case, the ratio between 
the PIMD displacement and force estimators,
$\gamma_{\textrm{\tiny PIMD}} \equiv \omega_{\delta x \delta x}/\omega_{FF}$,
gives exactly one. In the anharmonic situations instead, $\gamma_{\textrm{\tiny PIMD}}$ will deviate from the identity and it will be a quantitative signature 
of the anharmonic strength in the potential.
In the following, we will compare the value of $\gamma_{\textrm{\tiny PIMD}}$ with $\gamma_{exact} \equiv \frac{\omega_{1,0}}{\omega_0}$, i.e. the exact ratio between the first transition energy
and twice the ground state energy computed from the Schrödinger equation.
For the other anharmonic potentials, in order to test the PIMD estimators, we prefer to use a 
temperature as low as 20 K, where quantum effects are dominant.

\subsubsection{Morse potential}
The anharmonicity of the Morse potential is controlled by the $a_m$ parameter; in the limit of $a_m \to 0$, the potential becomes harmonic, while for increasing $a_m$, the potential's shape becomes more and more anharmonic.    
\begin{table}[h!]
\caption{Morse potential. $\omega_0$ is the ground state energy, $\omega_{1,0}$ the energy difference between the first excited state and the ground state. All the estimators are computed at 20 K and are given in cm$^{-1}$. $\gamma_{exact}$ is the ratio between $\omega_{1,0}$ and $\omega_0$.}
\begin{ruledtabular}
\begin{tabular*}{\textwidth}{{ccccc}}\label{tab:morse}

     {\small $\boldsymbol{a_m }$} & {\small $\boldsymbol{0.2 }$} & {\small $\boldsymbol{0.4 }$}  & {\small $\boldsymbol{0.6 }$} & {\small $\boldsymbol{0.8}$} \\
\hline
 $\omega_{FF}$ & 2193.55 & 2189.96 & 2183.96 & 2175.55 \\
 $\omega_0$ & 2193.54 & 2189.96 & 2183.98 & 2175.62 \\ \\
 $\omega_{\delta x \delta x}$ & 2190.86  &  2179.22 & 2159.82 & 2132.70 \\
 $\omega_{1,0}$ & 2189.97 & 2175.63 & 2151.75 & 2118.30 \\ \\
  $\gamma_{\textrm{\tiny PIMD}}$ &  0.9987 & 0.9950 & 0.9889 & 0.9803 \\
 $\gamma_{exact}$ & 0.9983 & 0.9934 & 0.9852 & 0.9736
\end{tabular*}
\end{ruledtabular}
\end{table}
From the results in Tab.~\ref{tab:morse}, we can observe that the estimator $\omega_{\delta x \delta x}$ 
reproduces the energy difference between the first excited state and the ground state,
as expected from Eq.~(\ref{eq:1d_expl}).
On the other hand, the force-force estimator closely follows the 
fundamental frequency value.
The ratio between the displacement-displacement and force-force estimators follows the trend of the true $\gamma_{exact}$ and, as expected, by increasing $a_m$ it deviates more and more from 1.

\subsubsection{Quartic potential}
At variance with Morse, the quartic potential cannot be reduced to the harmonic case by tuning the anharmonicity parameter $c_q$. 
The displacement-displacement estimator reproduces very well the energy difference between the first excited state and the ground state (Tab.~\ref{tab:quartic}) for all the three cases analysed. 
\begin{table}[h!]
\caption{Quartic potential. The notation is the same as in Tab.~\ref{tab:morse}.}
\begin{ruledtabular}
\begin{tabular*}{\textwidth}{{cccc}}\label{tab:quartic}

     {\small $\boldsymbol{c_q}$} & {\small $\boldsymbol{0.01 }$} & {\small $\boldsymbol{0.1 }$} & {\small $\boldsymbol{1.0}$} \\
\hline
 $\omega_{FF}$ & 332.74  & 716.86 & 1544.44 \\
 $\omega_0$ & 239.39 & 515.76 & 1111.17  \\ \\
  $\omega_{\delta x \delta x}$ & 310.85 & 669.71 & 1442.85 \\
  $\omega_{1,0}$ & 309.23 & 666.20 & 1435.30 \\ \\
 $\gamma_{\textrm{\tiny PIMD}}$ & 0.9342 & 0.9342 & 0.9342 \\
$\gamma_{exact}$ & 1.2917 & 1.2917 & 1.2917  
\end{tabular*}
\end{ruledtabular}
\end{table}

However,
$\omega_{FF}$ is not able to reproduce exactly the value of $\omega_0$. 
To investigate the reasons of such a mismatch, we compared $\omega_{FF}$ with the SCHA frequency,
defined as the best
harmonic approximation to the ground state energy, based on a variational principle for the free energy of the system\cite{Bianco_2017}. We found that the two values are very close each other, while being different from the exact ground state energy. This points to the fact that, for a quartic potential, the functional shape of the true ground state wave function is far from being harmonic, therefore invalidating the definition of the force constant matrix in Eq.~(\ref{eq:force_constant_matrix}) as a mean to recover the right fundamental frequency in this highly anharmonic case.
Due to the difference between $\omega_{FF}$ and $\omega_0$, we find that $\gamma_{\textrm{\tiny PIMD}}$ is always smaller than 1, while $\gamma_{exact}$ is larger. The quartic potential points out to the fact that the parameter $\gamma_{\textrm{\tiny PIMD}}$ cannot carry the information about the eventual superharmonic behavior of the potential energy (i.e. the case in which $\omega_{1,0}> \omega_0$). 
Nevertheless, its departure from 1, as reported in Tab.~\ref{tab:quartic}, still signals an anharmonic behavior, and
denotes correctly a larger deviation from harmonicity, if compared with the Morse potential.

\subsubsection{Symmetric double well potential}
The symmetric double well potential is physically relevant in situations where a hydrogen atom is shared by other two atoms to form a symmetric H-bond configuration. This occurs in many ferroelectrics or antiferroelectrics, where the hydrogen atom is shared by two oxygen or other electronegative atoms \cite{Matsushita_1982}. This also happens in the $H_5 0_2^+$ zundel cation \cite{Yu_2016,Mouhat_2017,Agostini_2011}.
By varying the $c_0$ parameter
of the symmetric double well potential,
we can range from the scenario where the barrier is smaller than the ground state energy ($c_0=0.05$ in Tab.~\ref{tab:sdw}) until the situation in which the barrier is much higher than the ground state ($c_0=0.3$ in Tab.~\ref{tab:sdw}). By analysing the results of Tab.~\ref{tab:sdw}, we observe again that $\omega_{\delta x \delta x}$ closely follows the $\omega_{1,0}$ value also in the extreme case in which the first excited state is close to the ground state ($c_0=0.3$). We deduce that the eigenvalue returned by the displacement-displacement estimator is reliable for evaluating the excitation energy spectrum over a wide range of anharmonic strengths.
The force-force frequency, instead, overestimates $\omega_0$ for c$_0 \leq 0.1$, while underestimates it when the barrier is higher.
\begin{table}[h!]
\caption{Symmetric double well potential. The notation is the same as in Tab.~\ref{tab:morse}. 
}
\begin{ruledtabular}
\begin{tabular*}{\textwidth}{{cccc}}\label{tab:sdw}

         {\small $\boldsymbol{c_{0} }$} & {\small $\boldsymbol{0.05 }$} & {\small $\boldsymbol{0.1 }$} & {\small $\boldsymbol{0.3}$} \\
\hline
 $\omega_{FF}$ & 1384.79 & 1321.63 & 2946.93 \\ 
 $\omega_0$ & 947.88 & 1100.10 & 3102.14 \\ \\
 $\omega_{\delta x \delta x}$ & 1190.52 & 925.44 & 124.85 \\ 
 $\omega_{1,0}$ & 1178.02 & 904.02 & 59.99  \\ \\
 $\gamma_{\textrm{\tiny PIMD}}$ & 0.8597 & 0.7002  & 0.0423 \\
 $\gamma_{exact}$ & 1.2427 & 0.8217  & 0.0193 
\end{tabular*}
\end{ruledtabular}
\end{table}
This can be rationalized with the fact that when c$_0$ is small, the behaviour of the symmetric double well is similar to the quartic potential case seen previously. On the other hand, when c$_0$ is higher, $\omega_{FF}$ tends to describe the curvature of one of the two wells. 

The extreme case of anharmonicity due to double well potentials with high barriers is nicely detected by
the ratio between the two quantum estimators $\gamma_{\textrm{PIMD}}$,
which shows a dramatic deviation from 1, in a very good agreement with $\gamma_{exact}$.

\subsubsection{Hydrogen molecule}
As final test for 1D potentials, we consider the realistic case of hydrogen molecule in the relative radial coordinate. 
We employed the effective potential from Ref.~\onlinecite{Kolos_1965} that reproduces very well (with an error of $\pm 1 $ cm$^{-1}$) the experimental Q$_1$(0) Raman lines \cite{Wolniewicz_1995,Allin_1965}.
\begin{table}[h!]
\caption{\label{tab:h2}%
Hydrogen molecule (here the potential minimum is set to zero). The notation is the same as Tab.~\ref{tab:morse}.}
\begin{tabular}{lr}
\colrule
\colrule
$\omega_{FF}$ \ \ \  &  4392.99  \\ 
$\omega_0$ \ \ \ &  4359.44 \\ 
 &  \\
$\omega_{\delta x \delta x}$ \ \ \ &  4218.38  \\
$\omega_{1,0}$ \ \ \  &  4162.08 \\ \\
$\gamma_{\textrm{\tiny PIMD}}$ &  0.9602 \\
$\gamma_{exact}$ & 0.9547 \\ 
\colrule
\colrule
\end{tabular}
\end{table}
As the hydrogen molecule potential in the radial coordinate looks like a Morse potential (with $a_m \sim 1.1$ using our notation), it is not surprising that we find the same behaviour of the estimators as in the Morse case. 
The same numerical value for $\omega_{\delta x \delta x}$ 
has been found and 
reported
in Ref.~\onlinecite{Ramirez_2001}. 

\subsection{Two-dimensional model}\label{sec:results_2d}
For the two-degree of freedom case, the Hamiltonian is given by
\begin{equation}
    H=\frac{p_1^2}{2m}+\frac{p_2^2}{2m}+V(x_1,x_2). \\
\end{equation}
To check the validity of the scheme to compute the 2D numerical Schrödinger equation, we analyse firstly two harmonic oscillators, i.e. $V(x_1,x_2)=k\cdot (x_1^2+x_2^2)$, coupled by a $\zeta \cdot x_1 \cdot x_2$ potential that can be solved analytically also in the context of PIMD \cite{Rossi_2017} or by perturbation theory for $\zeta \ll 1$. Results of this case are reported in appendix \ref{app:1d_2d_sch}.
\begin{figure}[h!]
\includegraphics[width=0.5\textwidth]{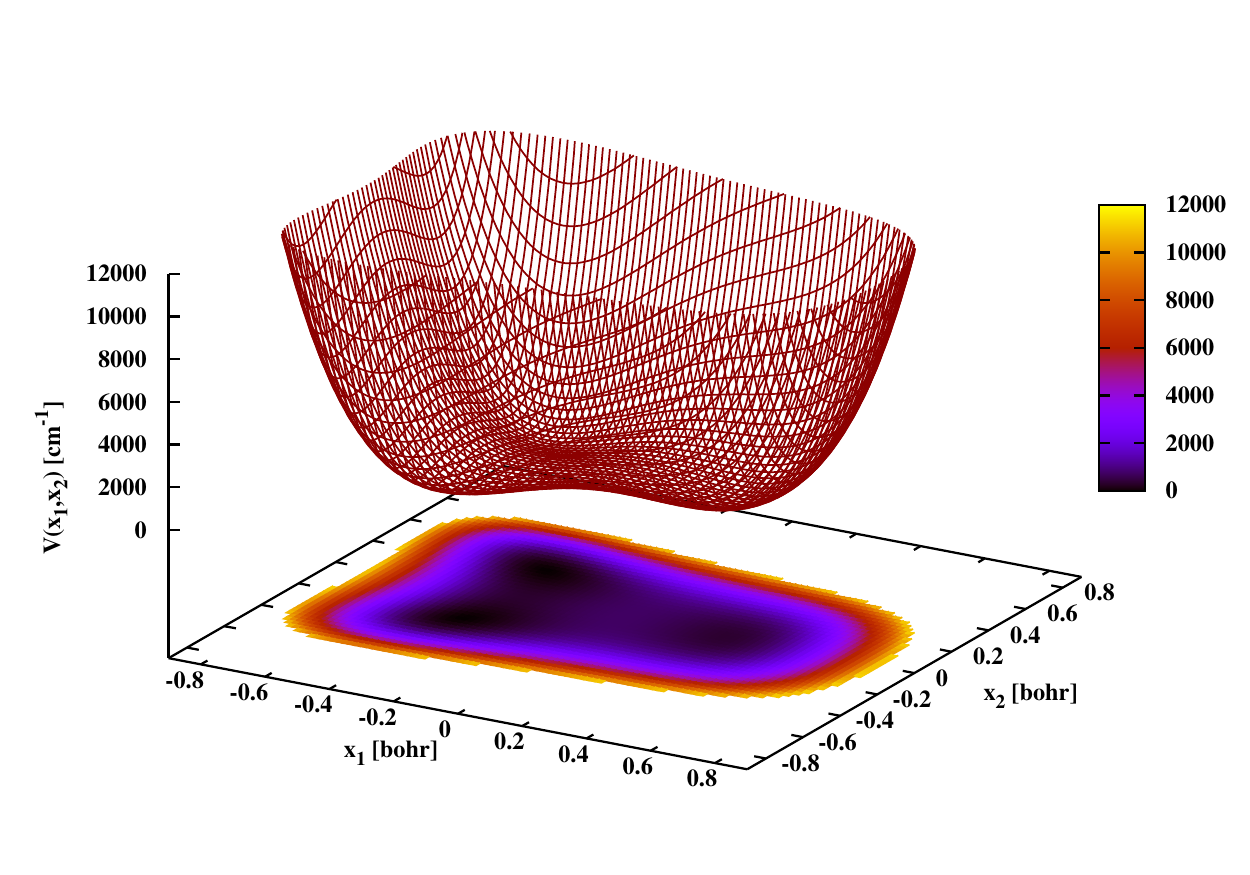}
\caption{\label{fig:sdw_qu_xy2} Symmetric double well potentials ($x_1$) plus quartic potential ($x_2$) coupled by a potential term of $x_1 \cdot x_2^2$ type.}
\end{figure}
Of course, in the 2D case, for each estimator we get two eigenvalues. Therefore, for clarity here we adopt the following notation: $\omega_{\delta x \delta x}$ and $\omega_{FF}$ acquire an index and become $\omega_{\delta x \delta x,i}$ and $\omega_{FF,i}$, where $i=1,2$. Moreover, the ground state energy, that we call $ZPE_{exact}$, is the sum of the zero point energies of the two modes. Thus, we estimate $ZPE_{exact}$ with $ZPE_{\textrm{PIMD}}=(\omega_{FF,1}+\omega_{FF,2})/2$.

A less trivial case is
a double well coupled with a quartic potential by a 
$\zeta \cdot x_1 \cdot x_2^2 $ term:
\begin{eqnarray}
    V(x_1,x_2) = k \cdot \left( x_1^2 - c_0 \right)^2 + 4k \cdot x_2^4 + \zeta \cdot x_1 \cdot x_2^2.
\end{eqnarray}
As in the 1D case, we choose $k=0.183736$ a.u.; then, we set $c_0=0.1$, and we varied $\zeta$ to switch from a non-interacting case ($\zeta=0$) to a strong interacting case ($\zeta=0.2$). 
In Fig.~\ref{fig:sdw_qu_xy2} we show the potential shape in the case of $\zeta=0.2$. In Tab.~\ref{tab:sdw_qu_xy2} we report the results. The situation in which $\zeta=0$ is the non-interacting case and the 2D problem reduces to two 1D problems. In the latter case, for the symmetric double well degree of freedom we get the same results as the 1D problem studied previously with $c_0=1$. By switching on the interaction, we observe that results agree with the findings in the 1D case. Indeed, despite 
the large
anharmonicity, 
one could expect 
only in 
some physical 
extreme 
conditions,
$\omega_{\delta x \delta x,1}$ and $\omega_{\delta x \delta x,2}$ are always very close to the first transition energies ($\omega_{1,0}$ and $\omega_{2,0}$ respectively) and represent an upper bound of them. On the other hand, the two frequencies given by the force-force estimator are harder to compare with the exact results because the $ZPE_{exact}$ value hides the contribution carried by the two modes. However, from the comparison between $ZPE_{\textrm{PIMD}}$ and $ZPE_{exact}$ we can conclude that, 
as previously found,
the fundamental frequency estimates 
become less accurate by increasing the interaction parameter $\zeta$.
\begin{table}[h!]
\small
\caption{Symmetric double well potential ($x_1$) coupled with a quartic potential ($x_2$) through a $\zeta \cdot x_1 \cdot x_2^2 $ interaction. We named the two modes for each estimator with (i=1) and (i=2), $ZPE_{\textrm{PIMD}}$ is the PIMD zero point energy computed as ($\omega_{FF,1}+\omega_{FF,2}$)/2  while $ZPE_{exact}$ is the ground state energy.}
\begin{ruledtabular}
\begin{tabular*}{\textwidth}{{llll}}\label{tab:sdw_qu_xy2}

     {\small $\boldsymbol{\zeta}$} & {\small $\boldsymbol{0.0 }$} & {\small $\boldsymbol{0.1 }$}  & {\small $\boldsymbol{0.2 }$} \\
\colrule
\\
  $\omega_{FF,i}$  &  \textbf{\scriptsize (i=1)} 1321.66  &  \textbf{\scriptsize (i=1)} 1400.28  & \textbf{\scriptsize (i=1)} 1615.27   \\
 &  \textbf{\scriptsize (i=2)} 2449.55  &  \textbf{\scriptsize (i=2)} 2410.38  & \textbf{\scriptsize (i=2)} 2314.16 \\
 $ZPE_{\textrm{PIMD}}$ & 1885.60 & 1905.33 &  1964.71 \\
 $ZPE_{exact}$ & 1785.52 & 1756.87 & 1665.86 \\ \\
 
 $\omega_{\delta x \delta x,i}$  &  \textbf{\scriptsize (i=1)} 925.14  &  \textbf{\scriptsize (i=1)} 959.07 & \textbf{\scriptsize (i=1)} 1063.03 \\
 &  \textbf{\scriptsize (i=2)} 2289.19  &  \textbf{\scriptsize (i=2)} 2199.82  & \textbf{\scriptsize (i=2)}  1941.19  \\ 
 $\omega_{1,0}$ & 903.72 & 927.48 & 999.57   \\ 
 $\omega_{2,0}$ & 2277.20 & 2162.48 & 1867.59  \\ \\

 $\gamma_{\textrm{\tiny PIMD}}^{\textbf{\scriptsize (i=1)}}$ & 0.699 & 0.6849 & 0.6581 \\ 
 $\gamma_{\textrm{\tiny PIMD}}^{\textbf{\scriptsize (i=2)}}$ & 0.9345 & 0.9126 & 0.8388 

\end{tabular*}
\end{ruledtabular}
\end{table}
The 2D models evidence that the overall performance of the estimators is still quantitative valid in the presence of interaction between anharmonic degrees of freedom. Therefore, we expect their reliability also in the truly many-body situation of real solids, as we will show in the diamond benchmark system.

\subsection{Diamond}\label{sec:results_diamond}
\begin{figure*}
\includegraphics[width=0.45\textwidth]{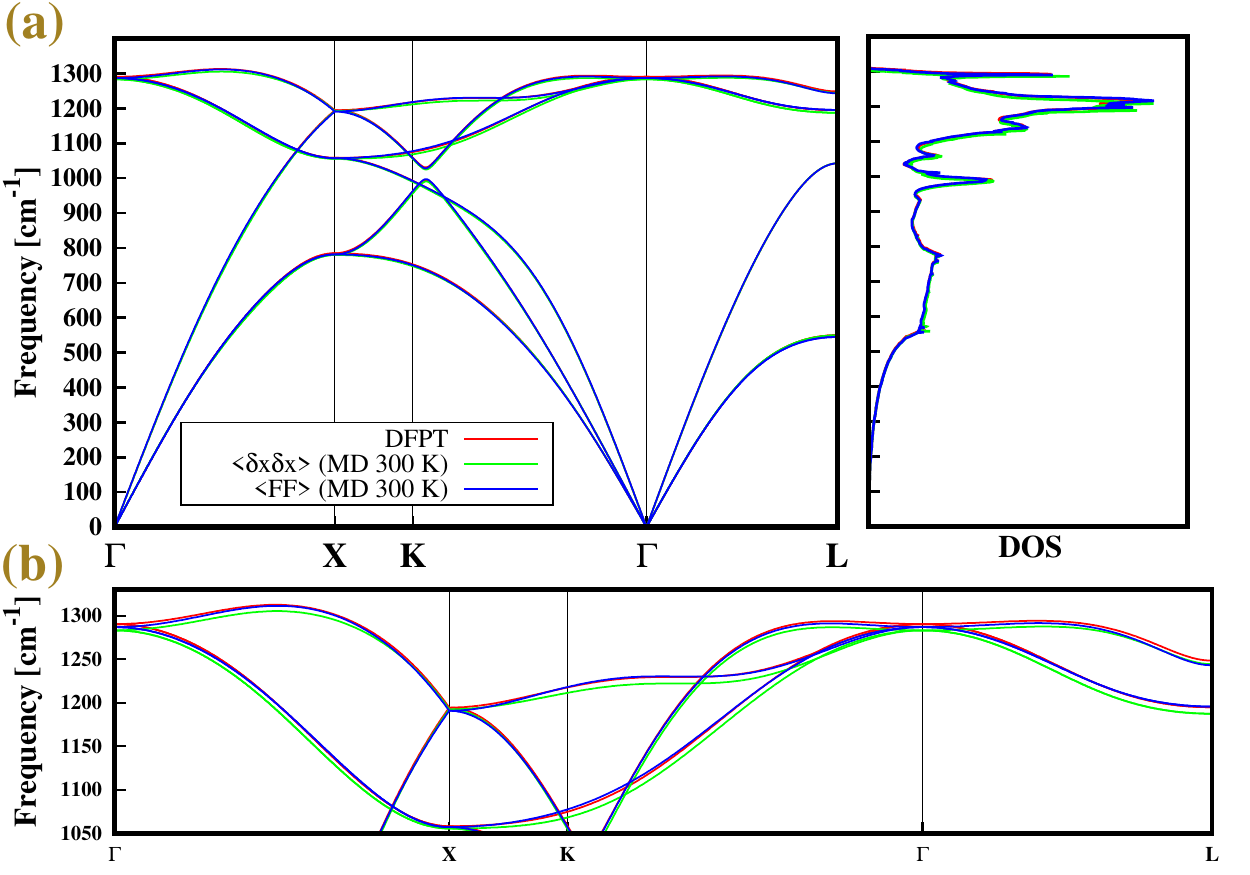}
\includegraphics[width=0.45\textwidth]{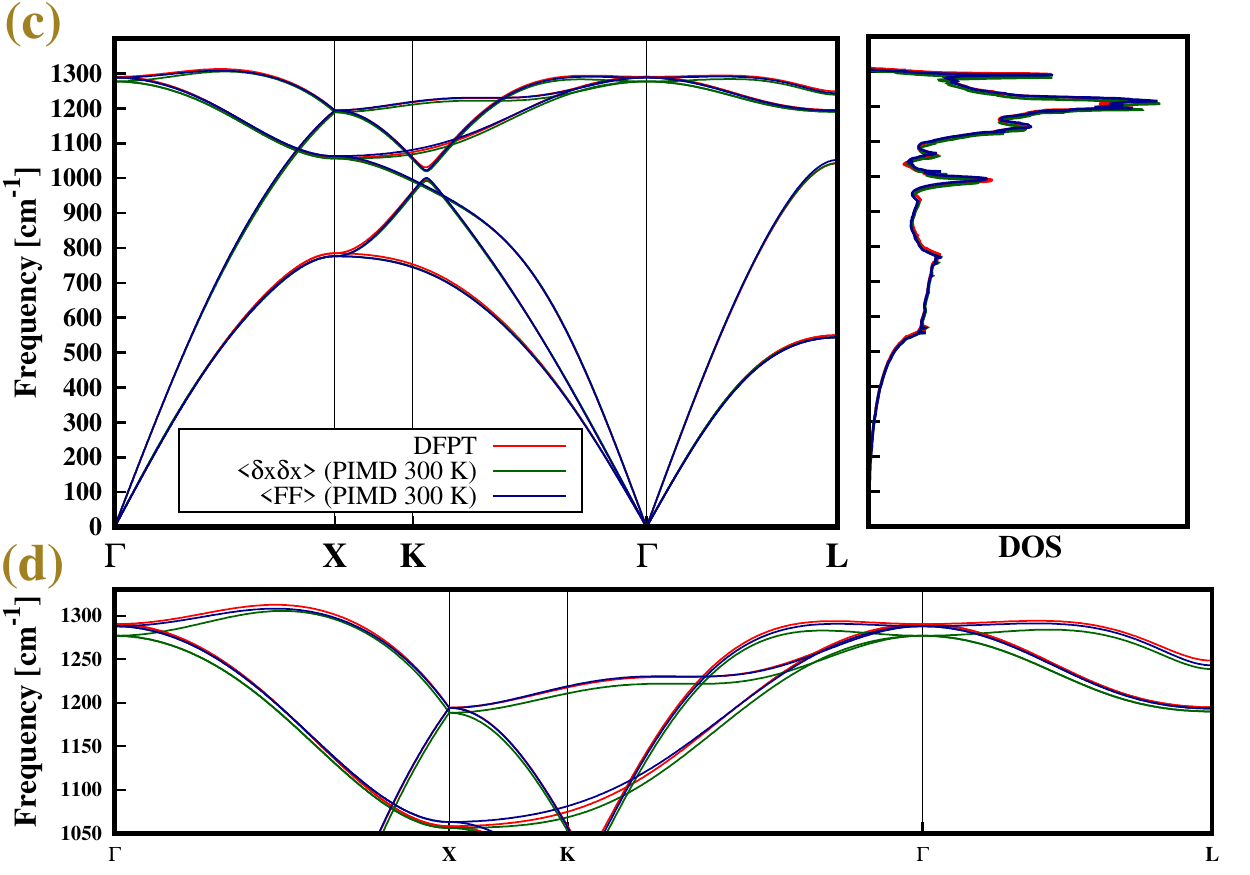}
\caption{\label{fig:diam} Diamond: (a) Phonon dispersion and DOS obtained from classical MD simulations at 300 K compared with DFPT calculation; (b) classical MD: zoom around the optical region; (c) Phonon dispersion and DOS obtained from PIMD simulations at 300 K with 12 beads compared with DFPT calculation; (d) PIMD: zoom around the optical region.}
\end{figure*}

The first crystalline system that we analyse 
is diamond, for which a vast
literature of both experimental and theoretical results exists \cite{Liu_2000,Kulda_2002,Schwoerer_1998,Pavone_1993}.
We performed a classical MD simulation and a PIMD one to compare the results and study the impact of quantum effects. In both cases, we kept the temperature constant at 300 K and in PIMD we employed 12 beads that were enough to converge the kinetic energy estimators.
For both MD and PIMD simulations, the supercell is made by $2 \times 2 \times 2$ conventional cells of diamond (each containing 8 atoms).
The convergence with respect to the supercell size and the other DFT parameters was checked using DFPT simulations as implemented within Quantum Espresso \cite{qe1} (QE). All the estimators in MD and PIMD were computed using a PBE exchange-correlation functional and ultrasoft pseudopotential that we took from the QE webpage.
The classical MD simulation was run using the algorithm implemented as in Ref.~\onlinecite{Ceriotti_2010} for the integration of EOMs. For this case, the time-step integrator was set to 0.5 fs, while DFT energy cut-off was set equal to 60 Ry for wavefunctions (480 Ry for charge density). The k-mesh to integrate over the Brillouin zone was equal to $2 \times 2 \times 2$ and the MD simulation lasted for 31 ps. Results are reported in Fig.~\ref{fig:diam}(a).

Diamond is an interesting system to test our estimators because at $\Gamma$ there is a weak but sizeable renormalization of the Raman frequency due to anharmonic effects, which is estimated to be 15 cm$^{-1}$ in Ref.~\onlinecite{Maezono_2007} and 17.4 cm$^{-1}$ in Ref.~\onlinecite{Vanderbilt_1984}. Raman frequencies are associated to transitions between vibrational (and rotational if present) states; therefore we expect to retrieve this effect through the quantum displacement-displacement estimator. 

The relative shift of the classical displacement-displacement with respect to the DFPT harmonic calculation at $\Gamma$ is 7$\pm$2 cm$^{-1}$ at 300 K. We exclude that the Raman shift is due to anharmonic temperature effects the classical phonon simulations could capture.  Indeed, 
it is known that the impact of temperature in diamond is marginal in the 
[0-300] K range.
Experimentally, the Raman mode shift is less than 1 cm$^{-1}$ in this temperature interval \cite{Liu_2000}. Thus, we deduce that this energy difference is due to anharmonic quantum effects that classical MD cannot fully capture. 
In order to recover a renormalization similar to the predicted one, one should take into account quantum effects. In Fig.~\ref{fig:diam}(c), we plot the phonon dispersions from PIMD simulations. The latter lasted 34 ps, with a time-step integrator of 0.75 fs. The DFT parameters for the calculation of the BO surface during the dynamics are the same as in the MD case. While the overall shape looks very similar to the classical MD case and to the DFPT calculation, in this case the displacement-displacement estimator gives us a renormalization of 13.7 $\pm$ 2 cm$^{-1}$ at $\Gamma$, in accordance with the theoretical prediction in Ref.~\onlinecite{Maezono_2007}.

Although we can reproduce the anharmonic shift at $\Gamma$ and the theoretical calculations at DFT-PBE level reported in Ref.~\onlinecite{Maezono_2007}, we notice that the absolute value of the optical mode at $\Gamma$ does not agree with the experimental one \cite{Kulda_2002,Liu_2000}. For instance, while the experimental first-order Raman mode is $\sim$1332.8 cm$^{-1}$, our displacement-displacement estimator gives 1276.9 cm$^{-1}$. This is due to the failure of the PBE functional that does not describe correctly the electronic correlation \cite{Maezono_2007}. To overcome this issue a full Quantum Monte Carlo calculation of the phonon dispersion for diamond was performed in Ref.~\onlinecite{Nakano_2020}. In particular, the $\langle \delta x \delta x \rangle $ curve in Fig.~\ref{fig:diam}(c), is the same reported in Ref.~\onlinecite{Nakano_2020} where it was used to renormalize the Quantum Monte Carlo phonon dispersion by anharmonic contributions. Using the Variational Monte Carlo as electronic solver, the estimated Raman frequency is 1336.9 cm$^{-1}$. 

\subsection{I4$_1/$amd high-pressure phase of hydrogen}\label{sec:results_hydrogen}
The determination of the phase diagram of solid hydrogen at high pressures is one of the major challenges in condensed matter physics. Indeed, as hydrogen is the lightest element,
it is
characterized by large NQE and show strong anharmonicity. 
\begin{figure*}
\includegraphics[width=0.48\textwidth]{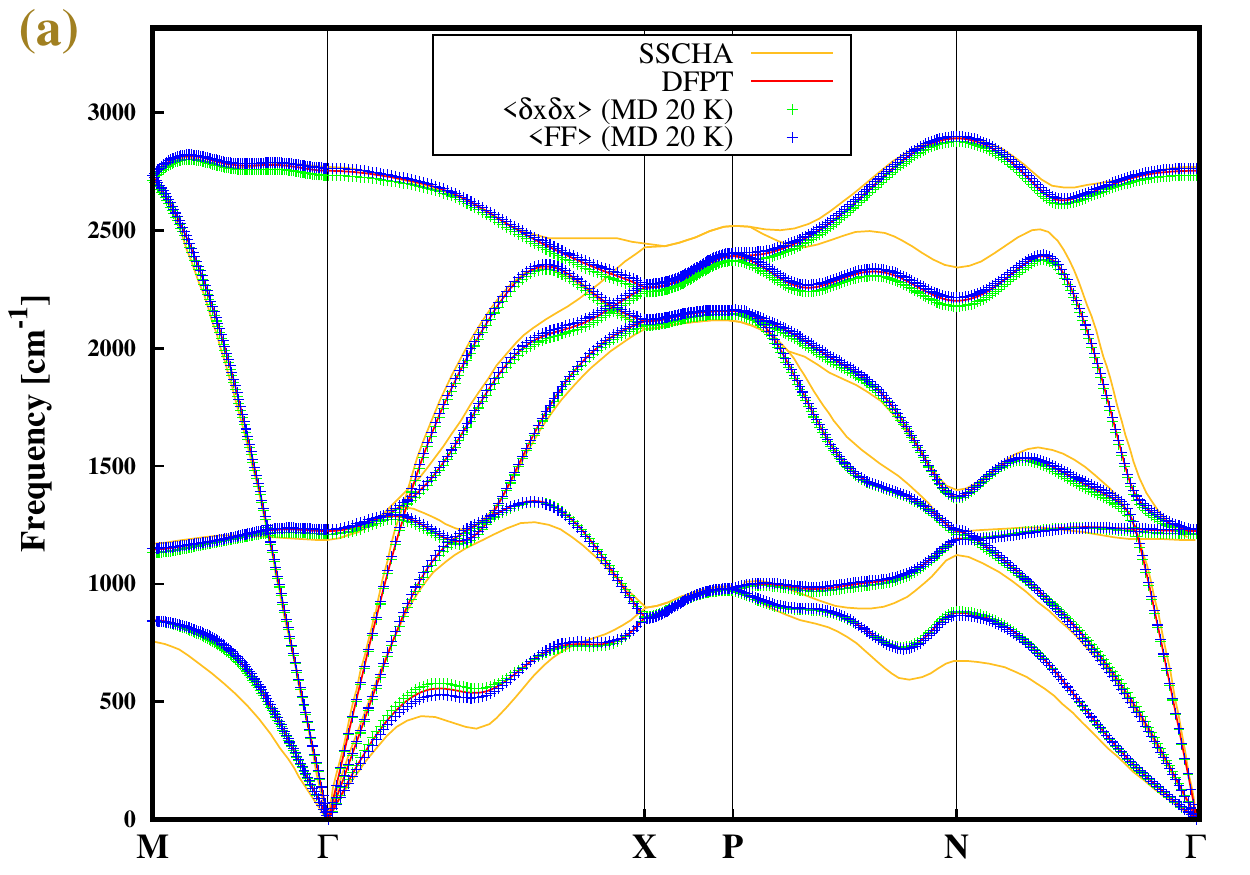}
\includegraphics[width=0.48\textwidth]{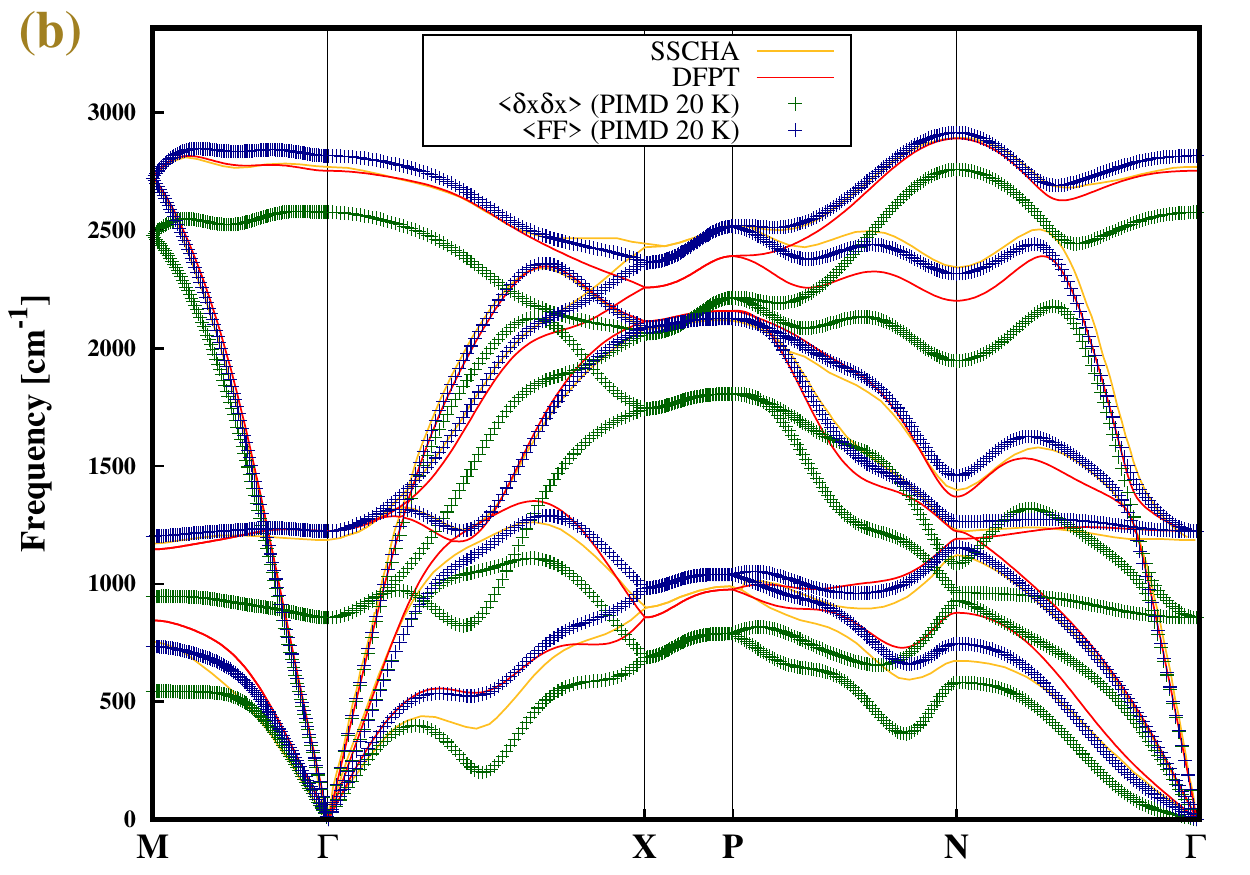}
\includegraphics[width=0.48\textwidth]{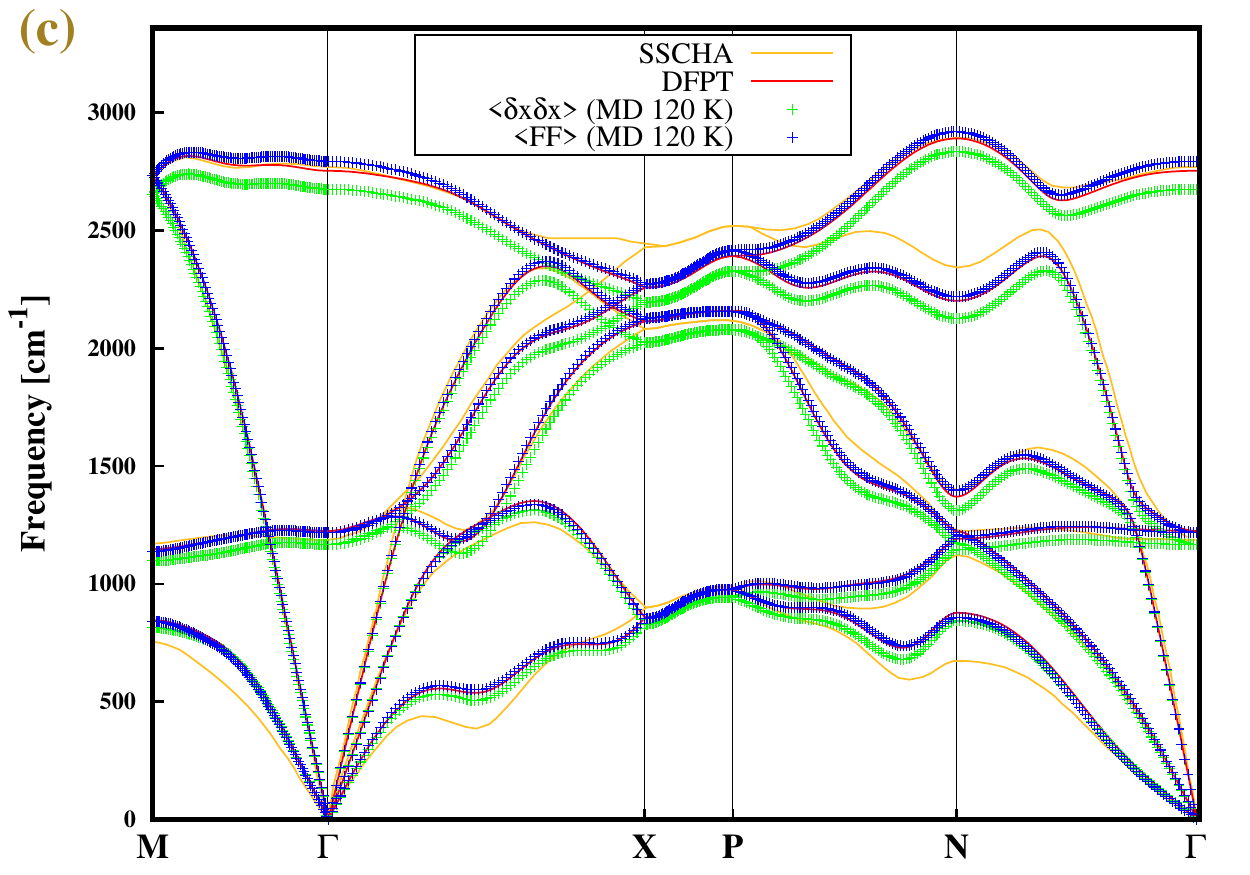}
\includegraphics[width=0.48\textwidth]{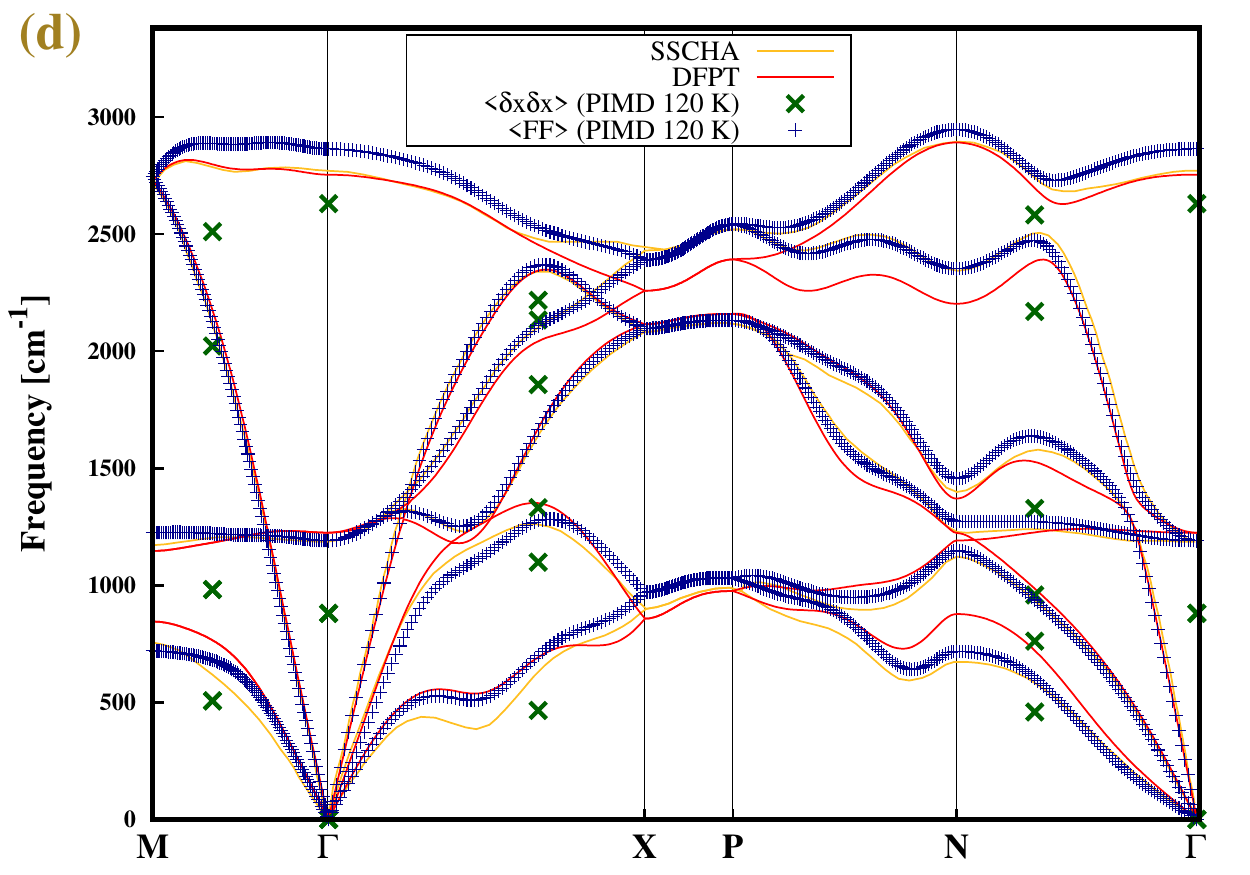}
\caption{\label{fig:hy_20K_120K} I4$_1$/amd atomic phase of hydrogen at 500 GPa: 
(a) Phonon dispersion obtained from MD simulations at 20 K compared with DFPT calculation (red) and SSCHA \cite{Borinaga_2016} (yellow). The red curve is perfectly covered by the force-force phonon estimator (blue); (b) Phonon dispersion obtained from PIMD simulations at 20 K compared with DFPT calculation (red) and SSCHA (yellow); (c) same as (a), but at 120 K; (d) same as (b) but at 120 K. In panel (d),
for the quantum displacement-displacement estimator, 
we plot the energies only at the q-points sampled by the simulations, because the dynamical matrix is harder to interpolate in this case. The data in Ref.~\onlinecite{Borinaga_2016} were digitalized using WebPlotDigitizer \cite{Rohatgi_2020}.}
\end{figure*}
Therefore, a good 
application
for the PIMD estimators is the I4$_1/$amd phase of atomic hydrogen at 500 Gpa, which is predicted to be the first atomic 
metallic phase above 490 Gpa \cite{Borinaga_2016,Pickard_2007}. 
We thus performed MD and PIMD simulations at both 20 K and 120 K, to compare the relative impact of quantum and thermal effects.

The unit cell of tetragonal I4$_1/$amd atomic hydrogen is 
made of
two atoms. We set $a=2.286$ bohr and $c=5.820$ bohr as lattice parameters, 
taken from
Ref.~\onlinecite{Borinaga_2016}.
These values correspond to a pressure of 500 GPa, as estimated by
DFT calculations with the
PBE exchange-correlation functional.
In order to sample the phonon dispersion, we used a $3 \times 3 \times 3$ supercell, made of 54 atoms. As this phase is metallic, a fine k-mesh in reciprocal space is required for the electronic integration
over the Brillouin zone (BZ). At DFPT level, 
convergence on the phonon dispersion is reached using a $30 \times 30 \times 30$ k-space grid in the unit cell.
This corresponds to a $10 \times 10 \times 10$ k-mesh for the supercell.
In our calculations, the plane-wave (density) energy cutoff is set equal to 50 Ry (330 Ry), 
and the Gaussian smearing is 
0.03 Ry.
\begin{figure*}
\includegraphics[width=0.48\textwidth]{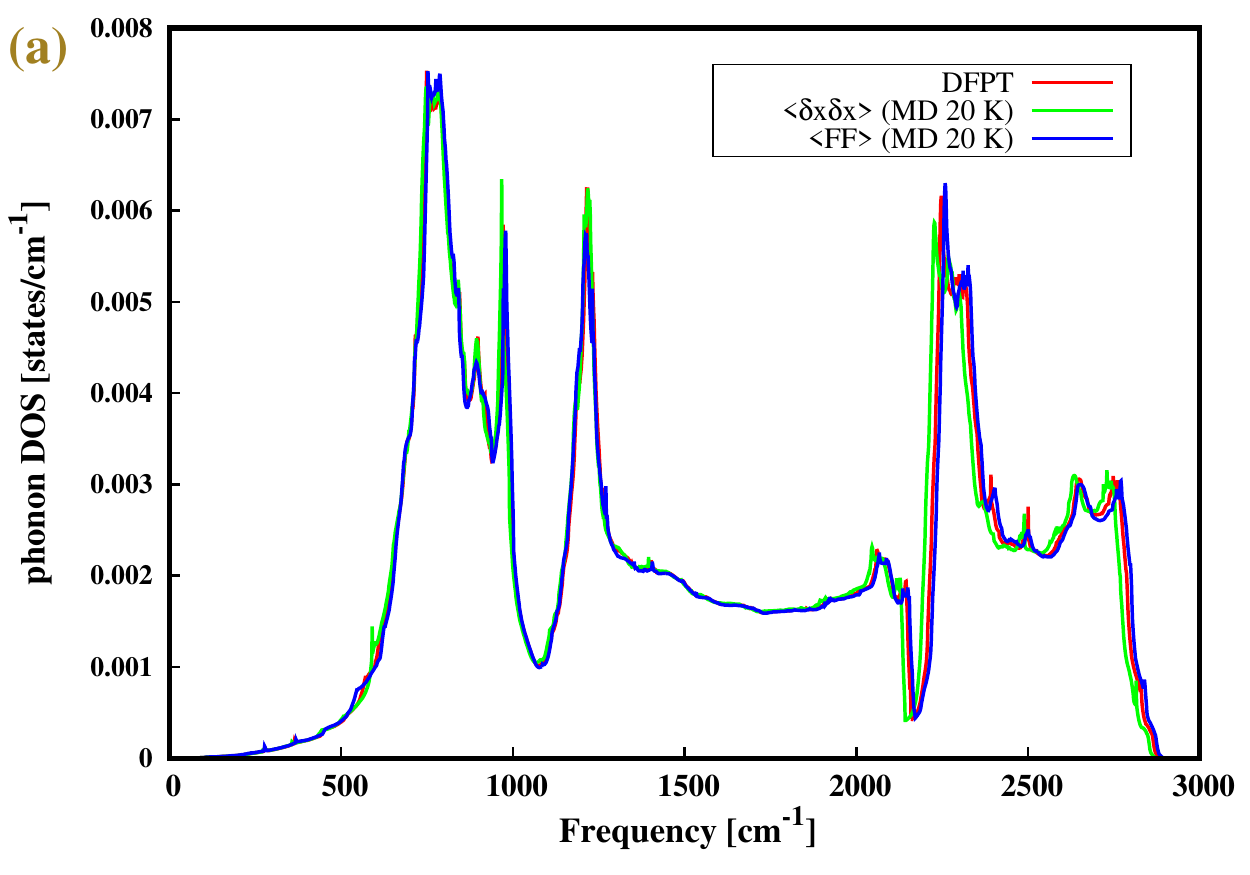}
\includegraphics[width=0.48\textwidth]{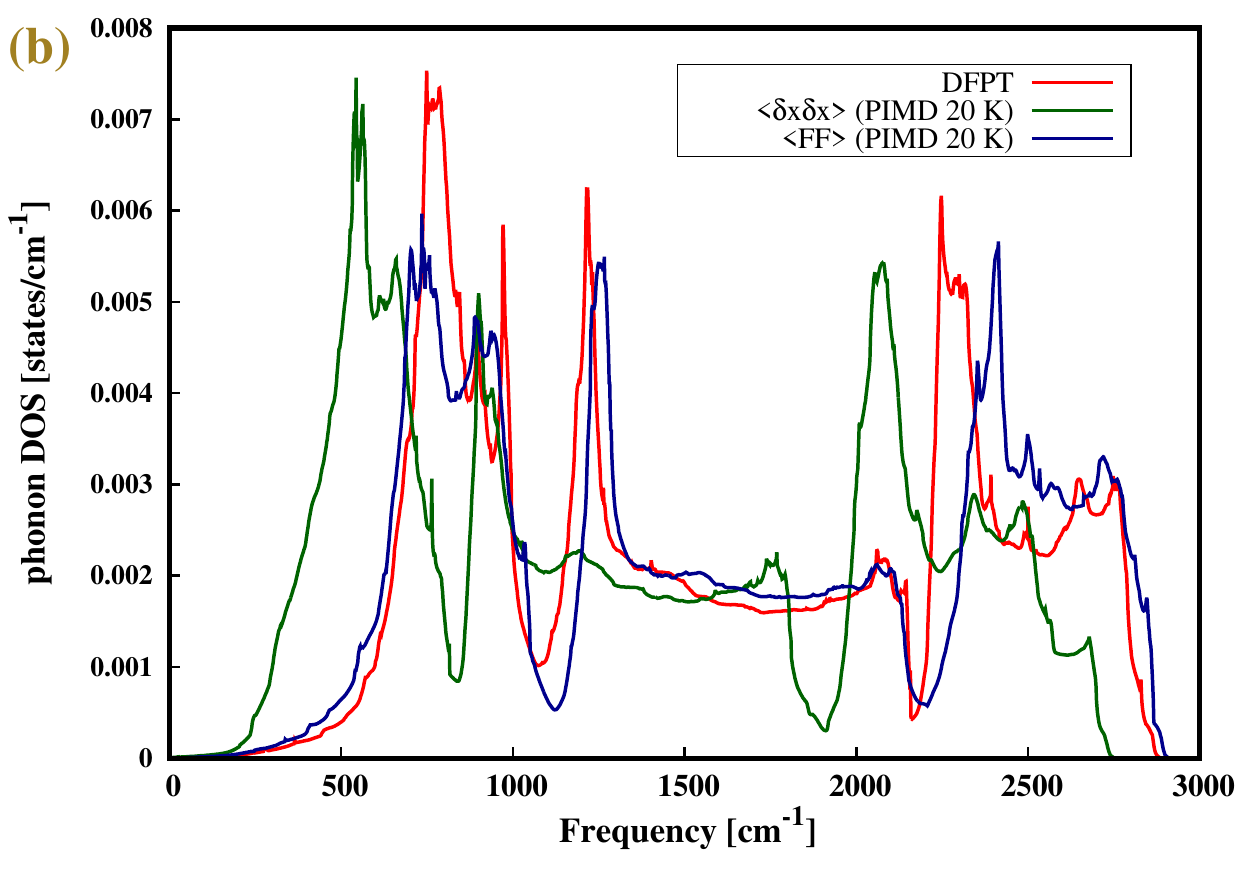}
\caption{\label{fig:hy_dos_20K} I4$_1$/amd atomic phase of hydrogen at 500 GPa: 
(a) Total phonon DOS obtained from classical MD simulations at 20 K compared with DFPT calculation (red); (b) Total phonon DOS obtained from PIMD simulations at 20 K compared with DFPT calculation (red).}
\end{figure*}
For both temperatures,
the classical MD time-step $\Delta t$ 
is 
0.5 fs , while in PIMD simulations $\Delta t = 0.9$
fs. In the latter case, the number of beads is $P=120$ ($P=50$) for the simulations at 20 K (120 K).
$P$ is chosen by studying the convergence of the virial and primitive kinetic energy estimators\cite{Ceriotti_2010,Mouhat_2017} (see appendix \ref{app:fcme_hy}).
The simulation length 
is 10 ps for classical MD, while it is 6 ps for PIMD. Remarkably, we find that 
such total simulation times
are enough to achieve an error bar of less than 10 cm$^{-1}$ on the phonon eigenvalues in almost all cases. This is thanks to the generalized eigenvalue problems, as formulated in Eqs.~(\ref{eq:ff_pimd}) and (\ref{eq:dxdx_pimd}). 
The hardest case is 
the PIMD simulation at 20 K, which requires a larger number of beads, $P=120$. Even in this situation, the error bar does not exceed 50 cm$^{-1}$.
In App.~\ref{app:fcme_hy}, we plot the convergence of phonon frequencies at $\Gamma$ and M points
as a function of the 
simulation time.
The frequencies are very stable after a short equilibration time of $\approx 1.5$ ps, with no drift and very small statistical fluctuations afterwards.

As a last technical remark, it is worth noting that 
the size of 
the $3 \times 3 \times 3$ q-point grid sampled by the MD/PIMD 
supercell is not enough to interpolate the phonon dispersion on a dense q-mesh. Indeed, from DFPT calculations, it turns out that a finer $6 \times 6 \times 6$ mesh is needed for interpolation.
Therefore, assuming that the anharmonic corrections are shorter-ranged, we interpolate the difference between the harmonic and anharmonic force constant matrices on the $3 \times 3 \times 3$ grid. We then add this anharmonic correction to the force constant matrix obtained on a finer $6 \times 6 \times 6$ mesh from DFPT calculations. This scheme has previously been employed in SSCHA calculations of analogous systems~\cite{Borinaga_2016,Errea_2013}.


In Figs.~\ref{fig:hy_20K_120K}(a) and \ref{fig:hy_20K_120K}(b), we plot the phonon dispersions at 20 K from classical and PIMD estimators, respectively; Figs.~\ref{fig:hy_20K_120K}(c) and \ref{fig:hy_20K_120K}(d) 
show
classical and quantum dispersions at 120 K. At 20 K, the classical MD estimators both agree very well with harmonic DFPT calculations, as one would expect 
at such a low temperature.
In the quantum case, the dispersion of the phonon frequencies computed through the force-force estimator follows well the SSCHA dispersion \cite{Borinaga_2016}, computed as q-dependent set of eigenvalues, diagonalizing the best auxiliary harmonic potential that minimizes the vibrational free energy. The force-force estimator is quite close to the harmonic DFPT phonons as well. However, the displacement-displacement PIMD estimator strongly renormalizes the harmonic curve. In particular, the quantum displacement-displacement optical branches are softened by 200-300 cm$^{-1}$ on average. 
Therefore, the simulations at 20 K reveal the presence of strong NQE and anharmonicity.
While we can quantify the anharmonic strength
by comparing the quantum force-force and displacement-displacement curves, in this case the temperature dependence of the classical simulations 
also suggests a large deviation from the harmonic behavior. Indeed,
at variance with the low temperature case, at 120 K the classical displacement-displacement estimator gives results that are significanlty different than DFPT (Fig.~\ref{fig:hy_20K_120K}(c)).  
Anharmonicity makes 
the classical phonon frequencies 
depend 
on temperature. On the other hand, the PIMD curves seem to be less affected by temperature effects, as NQE prevail (Fig.~\ref{fig:hy_20K_120K}(d)).

The shape of phonon dispersions is reflected in the phonon density of states (DOS), that we plot for classical MD and PIMD simulations at 20 K in Figs.~(\ref{fig:hy_dos_20K}(a)) and (\ref{fig:hy_dos_20K}(b)), respectively. In particular, the quantum displacement-displacement DOS can be directly compared with vibrational spectroscopies, as it is the 
density distribution
of the lowest-energy phonon excitations (green line in Fig.~(\ref{fig:hy_dos_20K}(b)). Instead, the force-force and harmonic DOS are 
ground state
approximations of the spectrum. Fig.~(\ref{fig:hy_dos_20K}(b)) shows that 
quantum
effects lead to a sizable red-shift,
due to anharmonicity, which amounts to $\approx 250$ cm$^{-1}$, if one compares lowest-energy excitations with fundamental frequencies.
A similar shift has been predicted in Ref.~\onlinecite{Monacelli_2020} for high-pressure molecular hydrogen, by means of a time-dependent SSCHA formulation.

From these results, we infer that the form of the potential felt by phonons in the atomic phase of hydrogen is far from being harmonic, as previously supposed \cite{Borinaga_2016} based on the estimate of the fundamental frequencies only. Rather, the comparison between 
fundamental
and 
excited-states
quantities 
obtained from the two quantum estimators allows one to capture a non-negligible anharmonic behavior. Thus, the consistent evaluation of both quantum estimators within the same PIMD framework is a valuable strategy to 
characterize the shape of the many-body interaction potential felt by the phonon quasiparticles.

\section{Conclusion}\label{sec:conclusion}
This work lays down  
the formalism to 
extend in a transparent way phonon dispersion calculations 
from classical to path integral molecular dynamics simulations. The proposed phonon quantum estimators are built upon zero-time Kubo-transformed quantum correlation functions, 
and are tested over toy-model potentials where exact results can be compared with from the numerical solution of the Schrödinger equation. 

We derived two classes of PIMD phonon estimators, based on the displacement auto-correlation and force auto-correlation functions, respectively. 
In particular, we have shown that while the force-force quantum correlators give access to the fundamental frequencies and thermodynamic properties of the quantum system,
the displacement-displacement correlators probe the lowest-energy phonon excitations, and so the low-energy phonon spectroscopy with high accuracy.
We have also shown that the simultaneous evaluation of quantum force-force and displacement-displacement phonon estimators gives a precious insight into the anharmonicity strength of the system,
within the same framework.

Furthermore, the rigorous quantization of classical equations performed in this work allowed us to write down a generalized eigenvalue problem for the effective normal modes determination, borrowed from the classical localization principle of the velocity power spectrum. 
We prove that the use of generalized eigenvalue equations in place of standard normal mode equations leads to a remarkable speed-up in the PIMD phonon calculations, both in terms of faster convergence rate and smaller time-step bias. 
The approach we propose gives converged eigenvalues at a lower computational cost and with a much smaller error bar. Overall, a one-order-of-magnitude gain is expected when PIMD phonons are computed through a generalized eigenvalues problem, either based on force or on displacement autocorrelation functions.

Our method relies on the PIOUD algorithm to integrate the path integral equation of motions in a Langevin framework. This algorithm allows 
one to run 
very efficient PIMD
simulations.
Our approach is fully \emph{ab initio}, the BO interaction potential for nuclei being computed at the DFT level. We applied this to the calculation of phonon properties of diamond and atomic hydrogen using PBE-DFT as solver for the electronic part. 
Nevertheless, concerning the BO surface, PIOUD 
can also deal with stochastic approaches such as Quantum Monte Carlo (QMC). Therefore, the PIMD formalism developed here for phonons calculations in quantum systems is already 
QMC-compliant \cite{Mouhat_2017,Rillo_2018}.
This feature can open interesting avenues in the future, and it represents a 
strong appeal
for using our method when facing with strongly correlated materials. 

Finally, 
we believe that 
this framework 
is
particularly suitable 
for a better understanding of the recent experimental discovery of high-temperature high-pressure superconductors,
such as H$_3$S and LaH$_{10}$, where the fundamental role played by hydrogen in making these materials superconducting is already known \cite{Errea_2016,Errea_2020}. Indeed, our approach takes into account NQE, phonon anharmonicity and temperature effects together, in a non-perturbative way, without relying on any 
harmonic or self-consistent theory. 
In this sense,
it goes beyond 
current
approximations 
limiting
state-of-the-art methods.

\begin{acknowledgments}
T.M. and M.C. are grateful to the French grand {\'e}quipement national de calcul intensif (GENCI) for the computational time provided 
under Project No.~0906493.
T.M. and M.C. acknowledge that this work was supported by French state funds managed by the ANR within the Investissements d'Avenir programme under reference  ANR-11-IDEX-0004-02, and more specifically within the framework of the Cluster of Excellence MATISSE led by Sorbonne University.
All the authors thank Lorenzo Monacelli for useful discussions.

This work is partially supported by the European Centre of Excellence in Exascale Computing TREX - Targeting Real Chemical Accuracy at the Exascale. This project has received funding from the European Union’s Horizon 2020 - Research and Innovation program - under grant agreement no. 952165.
\end{acknowledgments}

\section*{Data Availability}
The data that support the findings of this study are available from the corresponding author upon reasonable request.

\appendix

\section{Relation between Kubo-transformed force constant matrix and PIMD force-force correlation functions }\label{app:dxdx}
In this appendix we review some properties of the PIMD force-force correlation functions. 

The Kubo-transformed force constant matrix can be derived exactly from the force-force correlation matrix in the following way:
\begin{eqnarray}\label{eq:average_ff_pimd_expl}
&& \llrrangle{\frac{1}{P^2}  \sum_{j_1,j_2=1}^{P} \frac{\partial^2 \mathsf{V}}{\partial x^{(j_1)}_{i_1} \partial x^{(j_2)}_{i_2}}} 
= \nonumber\\
&&=\int d^f \mathbf{p} \ e^{- \beta_P T_{P}} \int d^f \mathbf{x} \ \frac{1}{P^2}\sum_{j_1,j_2=1}^{P} \frac{\partial^2 \mathsf{V}}{\partial x^{(j_1)}_{i_1} \partial x^{(j_2)}_{i_2}} \frac{e^{- \beta_P V_{P}}}{Z} \nonumber\\
&&= \beta_P \int d^f \mathbf{p} \ e^{- \beta_P T_{P}} \int d^f \mathbf{x} \ \frac{1}{P^2} \Big( \sum_{j_1=1}^{P} \frac{\partial \mathsf{V}}{\partial x^{(j_1)}_{i_1}} \Big) \Big( \sum_{j_2=1}^{P} \frac{\partial V_{P}}{\partial x^{(j_2)}_{i_2}} \Big)  \frac{e^{- \beta_P V_{P}}}{Z} \nonumber\\
&&= \beta \int d^f \mathbf{p} \ e^{- \beta_P T_{P}} \int d^f \mathbf{x} \ \frac{1}{P^2} \Big( \sum_{j_1=1}^{P} \frac{\partial \mathsf{V}}{\partial x^{(j_1)}_{i_1}} \Big) \Big( \sum_{j_2=1}^{P} \frac{\partial \mathsf{V}}{\partial x^{(j_2)}_{i_2}} \nonumber \\ 
&& \ \ \ \ \ \ \ \   + \frac{1}{P}\sum_{j_2=1}^{P} m_{i_2} \omega_P^2 (  2x^{(j_2)}_{i_2} - x^{(j_2+1)}_{i_2} - x^{(j_2-1)}_{i_2} ) \Big)  \frac{e^{- \beta_P V_{P}}}{Z} \nonumber\\
&&= \beta \int d^f \mathbf{p} \ e^{- \beta_P T_{P}} \int d^f \mathbf{x} \ \frac{1}{P^2} \Big( \sum_{j_1=1}^{P} \frac{\partial \mathsf{V}}{\partial x^{(j_1)}_{i_1}} \Big) \Big( \sum_{j_2=1}^{P} \frac{\partial \mathsf{V}}{\partial x^{(j_2)}_{i_2}} \Big)  \frac{e^{- \beta_P V_{P}}}{Z} \nonumber\\
&&= \beta \int d^f \mathbf{p} \ e^{- \beta_P T_{P}} \int d^f \mathbf{x} \ \mathsf{F}_{i_1} \mathsf{F}_{i_2}  \frac{e^{- \beta_P V_{P}}}{Z} = \ \ \beta \llrrangle{\mathsf{F}_{i_1} \mathsf{F}_{i_2}},
\end{eqnarray}
where we integrated by parts with respect to $x^{(j_2)}_{i_2}$ between the second and the third row and where we used the fact that $\sum_{j_2=1}^{P}(  2x^{(j_2)}_{i_2} - x^{(j_2+1)}_{i_2} - x^{(j_2-1)}_{i_2} )=0$.
This relation extends the classical one in Eq.~(\ref{eq:ff_class}) in a transparent way. On the other hand, if one considers the equal-time correlation functions in Eq.~(\ref{eq:pimd_AB}) and tries to obtain the force constant matrix using them instead of the Kubo ones, he would end up with:
\begin{eqnarray}\label{eq:average_ff_pimd_expl_eqtime}
&& \llrrangle{\frac{1}{P}  \sum_{j=1}^{P} \frac{\partial^2 \mathsf{V}}{\partial x^{(j)}_{i_1} \partial x^{(j)}_{i_2}}} 
= \nonumber\\
&&=\int d^f \mathbf{p} \ e^{- \beta_P T_{P}} \int d^f \mathbf{x} \ \frac{1}{P}\sum_{j=1}^{P} \frac{\partial^2 \mathsf{V}}{\partial x^{(j)}_{i_1} \partial x^{(j)}_{i_2}} \frac{e^{- \beta_P V_{P}}}{Z} \nonumber\\
&&= \beta_P \int d^f \mathbf{p} \ e^{- \beta_P T_{P}} \int d^f \mathbf{x} \ \frac{1}{P} \Big( \sum_{j=1}^{P} \frac{\partial \mathsf{V}}{\partial x^{(j)}_{i_1}} \frac{\partial V_{P}}{\partial x^{(j)}_{i_2}} \Big)  \frac{e^{- \beta_P V_{P}}}{Z}, 
\end{eqnarray}
where the term inside the parenthesis is 
\begin{eqnarray}
    && \sum_{j=1}^{P} \frac{\partial \mathsf{V}}{\partial x^{(j)}_{i_1}} \frac{\partial V_{P}}{\partial x^{(j)}_{i_2}}= \nonumber \\ && =\sum_{j=1}^{P} F^{(j)}_{i_1} \Big[ F^{(j)}_{i_2} +  m_{i_2} \omega_P^2 \left(  2x^{(j)}_{i_2} - x^{(j+1)}_{i_2} - x^{(j-1)}_{i_2} \right) \Big].
\end{eqnarray} 
Last expression clearly shows that the equal-time force correlation contains the fictitious interbead coupling terms which make it different from the physical force constant matrix. For our purposes, this result supports the use of Kubo-transformed correlation functions in place of the equal-time ones.  

\section{Low temperature limit of quantum correlation functions}\label{app:details_c}
In the limit of low temperature $T \to 0$ ($\beta \to \infty$ and $Z \approx \sum^{N_{deg}}_{i=1} e^{-\beta E_i}$ where $N_{deg}$ is the number of states close to the ground state, as in the double well potentials with high barriers), the true quantum thermal zero-time correlation function, decomposed like the Kubo-transformed one in Eq.~(\ref{eq:kubo_decomp}) becomes: 
\begin{eqnarray}
    \lim_{\beta \to \infty} c_{AB} &&= \frac{1}{Z} \sum^{N_{deg}}_{n=1} e^{-\beta E_n} \left[ A_{nn} B_{nn} + \sum_{m \ne n} A_{nm} B_{mn}  \right] \nonumber\\
    && \text{if $N_{deg}$=1} \ \ \ c_{AB} \sim  A_{00} B_{00} + \sum_{m > 0}  A_{0m} B_{m0}.
\end{eqnarray}
It is worth noting that if $N_{deg}$=1, at $t=0$ the true quantum correlation functions can be written as an expectation value over the ground state.
\begin{equation}\label{eq:tqu_nodeg_low_T}
    \lim_{\beta \to \infty} c_{AB} \sim \langle 0 \vert A \ B^+ \vert 0 \rangle.
\end{equation}
On the other hand for the Kubo-transformed correlation function one can find:

\begin{eqnarray}\label{eq:kubo_low_T}
    \lim_{\beta \to \infty}  \tilde{c}_{AB} &&= \frac{1}{Z} \sum^{N_{deg}}_{n=1} e^{-\beta E_n} \left[ A_{nn} B_{nn} + 2 \sum_{m > n} A_{nm} B_{mn} \cdot \frac{1-e^{-\beta \hbar \omega_{m,n}}}{\beta \cdot \hbar \omega_{m,n}} \right] \nonumber \\
    && \text{if $N_{deg}=1$} \ \ \ \tilde{c}_{AB} \sim A_{00} B_{00} + 2 \sum_{m > 0} \frac{ A_{0m} B_{m0}}{\beta \cdot \omega_{m,0}}.
\end{eqnarray}
Differently from the true quantum case, it is worth noting that the zero-time Kubo correlator cannot be reduced into an expectation value over the ground state as Eq.~(\ref{eq:tqu_nodeg_low_T}). 

\section{Convergence of the phonon frequencies from the generalized eigenvalue equation versus the standard eigenvalue equation}\label{app:gen_vs_sta}
Here we investigate the behaviour of the GEV and standard estimators already presented in the main text in \ref{sec:ff_dxdx_pimd} for a Morse potential. 
\begin{figure}[h!]
\includegraphics[width=0.5\textwidth]{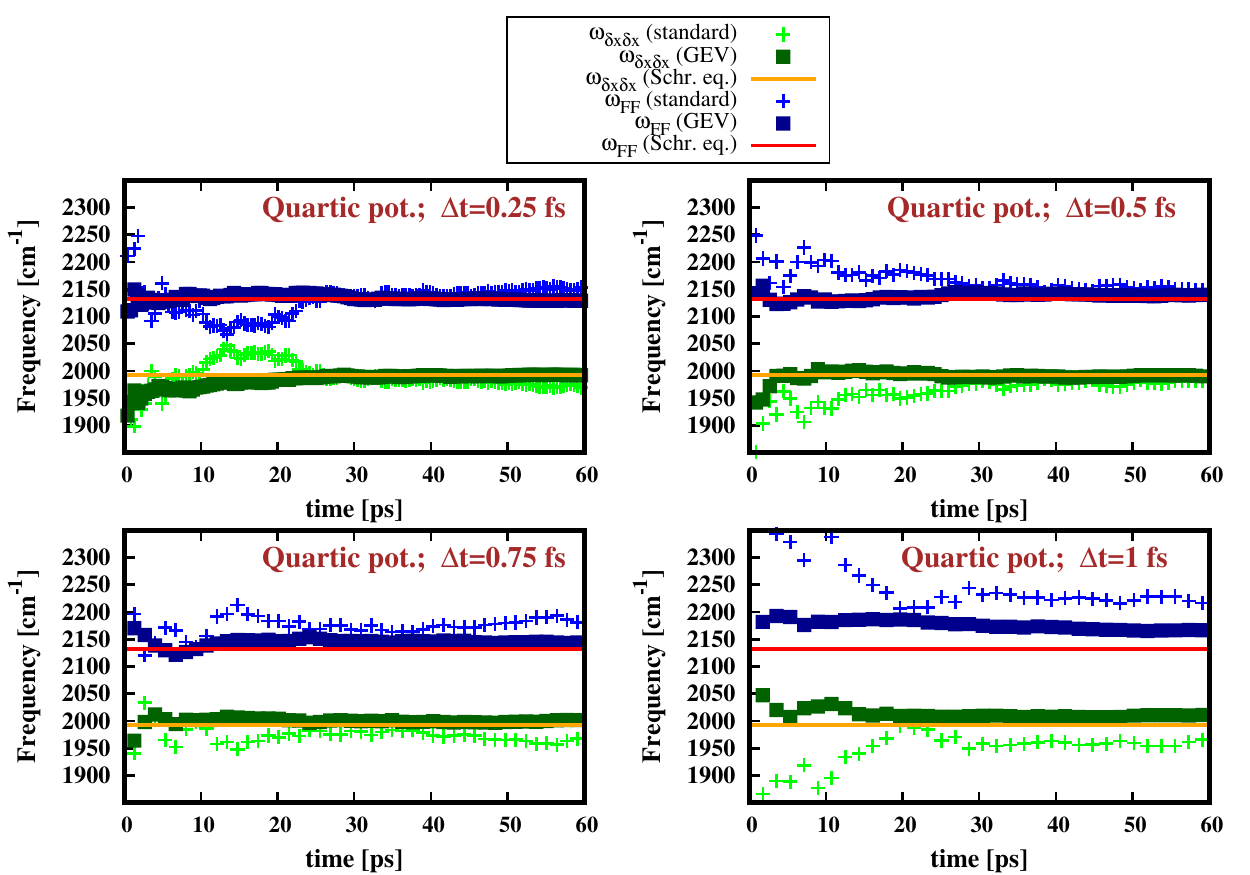}
\caption{\label{fig:ff_dxdx_qu_conv} Convergence of the PIMD estimators for a 1D degree of freedom bounded by a quartic potential at 100 K obtained with different time-step $\Delta t$.
The exact solutions that we obtain from the numerical Schrödinger equation are given in red for the force-force and orange for the displacement-displacement.}
\end{figure}
\begin{figure}[h!]
\includegraphics[width=0.5\textwidth]{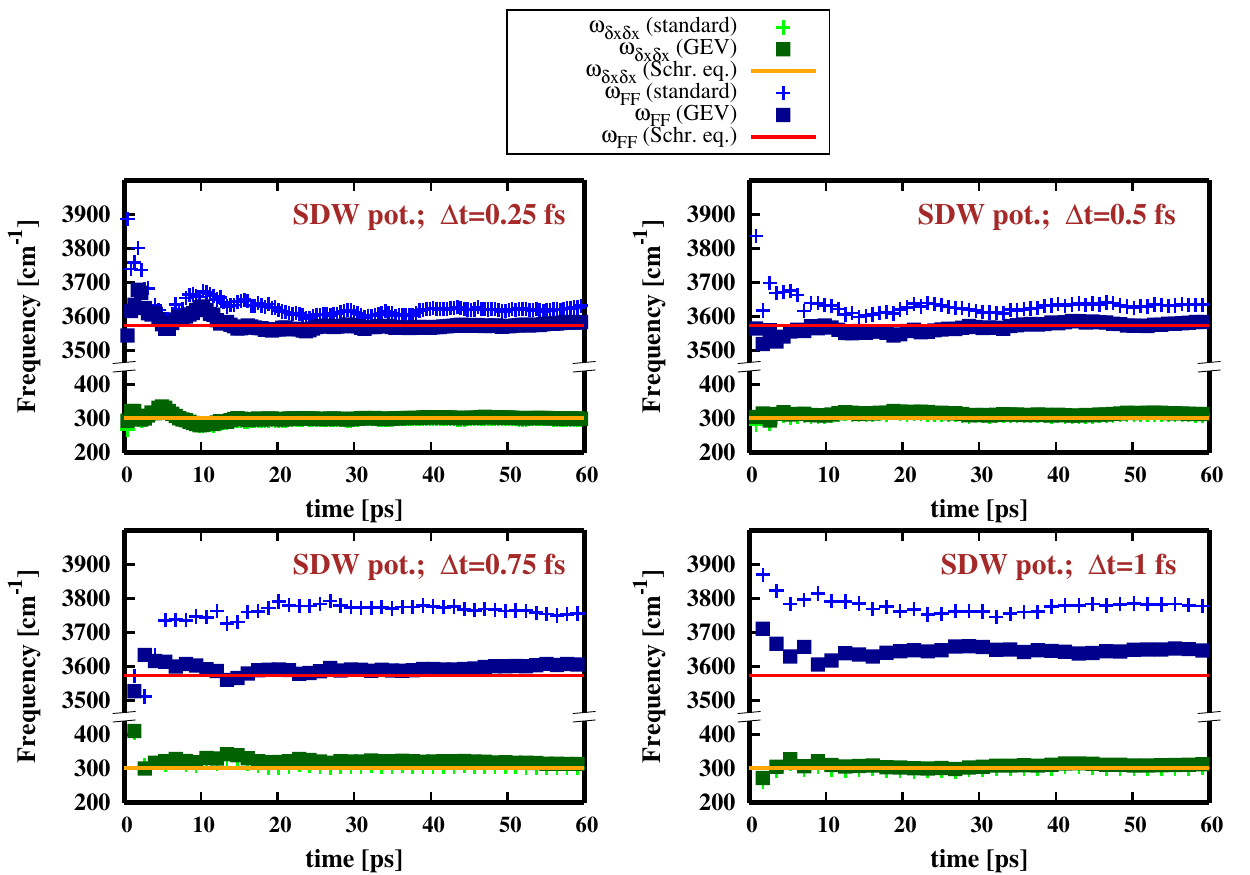}
\caption{\label{fig:ff_dxdx_sdw_conv} Convergence of the PIMD estimators for a 1D degree of freedom bounded by a symmetric double well potential at 100 K obtained with different time-step $\Delta t$. 
The exact solutions that we obtain from the numerical Schrödinger equation are given in red for the force-force and orange for the displacement-displacement.}
\end{figure}
We extend this analysis by including the case of a quartic potential (Fig.~(\ref{fig:ff_dxdx_qu_conv})), that we choose equal to $V(x)=k \cdot x^4$ and a symmetric double well potential (SDW in  Fig.~(\ref{fig:ff_dxdx_sdw_conv})), $V(x)=k\cdot(x^2-0.19)^2$, where $k=0.483736$ a.u. like in the Morse case. The temperature is always equal to 100 K and the number of beads is equal to 120 for the quartic and 160 for the symmetric double well. 
We observe that GEV solutions converge always faster and with less oscillations to the exact value (computed from the Schrödinger equation). However, at variance with the Morse potential reported in the main text, we denote a small error of the GEV solutions for a time-step integrator of 1 fs.


\section{Schemes for the solution of the numerical Schrödinger equations}\label{app:1d_2d_sch}
In the 1D case, the Schrödinger equation is discretized using finite differences on a fine mesh ($2 \cdot 10^5$ points). The discretized Hamiltonian is tridiagonal, thus easily diagonalizable using Lapack package \cite{lapack_1999}. Indeed, the potentials that we consider are bounded for $x \to \pm \infty$ except for the Morse that for $x \to +\infty$ tends to a finite value. However, in the region where $\frac{dV}{dx}=0$, we are able to reproduce analytical results for the harmonic and Morse potential eigenvalues with an error $< 0.01$ cm$^{-1}$ until the tenth bound state, that is enough to build our correlation functions at 20 K. 
\begin{figure}[h!]
\includegraphics[width=0.5\textwidth]{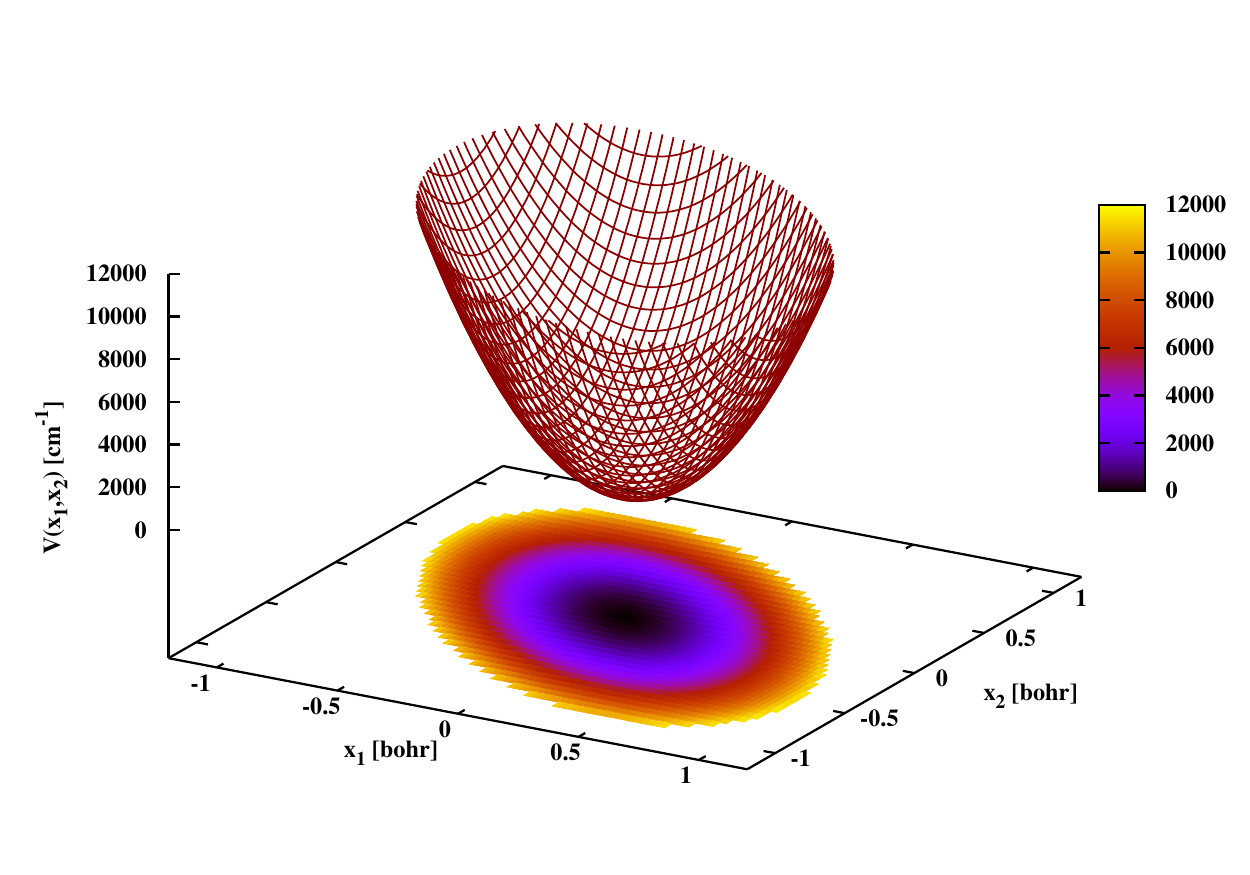}
\caption{\label{fig:ha_ha_2d} Two harmonic potentials coupled by a term of the type $\zeta \cdot x_1 \cdot x_2$.}
\end{figure}
For the 2D case, we still use finite differences for discretizing the Schrödinger equation in a real space square grid, but then the Laplace operator is not tridiagonal. Our real space grid is made of $1.5 \cdot 10^3 \times 1.5 \cdot 10^3$ points, resulting in a $2.25 \cdot 10^6 \times 2.25 \cdot 10^6$ Hamiltonian matrix.
\begin{table}[h!]
\caption{Two harmonic oscillators ($c_{11}=c_{12}=0.147$ a.u. in Eq.~(\ref{eq:ha_ha_2d})) with coupling potential $\zeta \cdot x_1 \cdot x_2$. The energies are given in cm$^{-1}$.}
\begin{ruledtabular}
\begin{tabular*}{\textwidth}{{llll}}\label{tab:ha_ha_2d}
     {\small $\boldsymbol{\zeta/c_{11} }$} & {\small $\boldsymbol{0.0 }$} & {\small $\boldsymbol{0.1 }$}  & {\small $\boldsymbol{0.5 }$} \\
\colrule
 $\omega_{FF,i}$ &  \textbf{\scriptsize (i=1)} 1963.29 &  \textbf{\scriptsize (i=1)} 1862.54 & \textbf{\scriptsize (i=1)} 1388.25  \\
 &  \textbf{\scriptsize (i=2)} 1963.29 &  \textbf{\scriptsize (i=2)} 2059.12  & \textbf{\scriptsize (i=2)} 2404.53 \\ 
 $ZPE_{\textrm{PIMD}}$ & 1963.29 & 1960.83 & 1896.39 \\
 $ZPE_{exact}$ & 1962.91 & 1960.45 & 1896.03 \\  \\

 $\omega_{\delta x \delta x,i}$ &  \textbf{\scriptsize (i=1)} 1962.78 &  \textbf{\scriptsize (i=1)} 1862.05 & \textbf{\scriptsize (i=1)} 1387.90  \\
 &  \textbf{\scriptsize (i=2)} 1962.78 &  \textbf{\scriptsize (i=2)} 2058.58 & \textbf{\scriptsize (i=2)} 2403.91 \\ 
 $\omega_{1,0}$ & 1962.78 & 1862.05 & 1387.90   \\ 
 $\omega_{2,0}$ & 1962.78 & 2058.58 & 2403.91  \\ \\

 $\gamma_{\textrm{\tiny PIMD}}^{\textbf{\scriptsize (i=1)}}$ & 0.9997 & 0.9997 & 0.9997  \\ 
 $\gamma_{\textrm{\tiny PIMD}}^{\textbf{\scriptsize (i=2)}}$ & 0.9997 & 0.9997 & 0.9997  
\end{tabular*}
\end{ruledtabular}
\end{table}
Instead of diagonalizing exactly the huge discretized Hamiltonian, we employ the Arnoldi package \cite{Lehoucq_1997} to compute only the first twenty eigenvalues and eigenvectors, that are enough to evaluate the estimators. The wider mesh grid gives an error of the order $< 2$ cm$^{-1}$ on the eigenvalues that we check comparing them with 2D harmonic oscillators analytical solutions.
In picture (\ref{fig:ha_ha_2d}) we show an example of 2D harmonic oscillators coupled by an $\zeta \cdot x_1 \cdot x_2$ term: 
\begin{equation}\label{eq:ha_ha_2d}
    V(x_1,x_2) = c_{11} \cdot x_1^2  + c_{12} \cdot x_2^2 + \zeta \cdot x_1 \cdot x_2.    
\end{equation}
While we know that in the harmonic case, both the displacement-displacement and force-force estimators give the same correct result, in Tab.~\ref{tab:ha_ha_2d} we can observe a difference always smaller than 1 cm$^{-1}$ between displacement-displacement and force-force which is due to numerical errors.

\section{Convergence of phonon frequencies for I4$_1$/amd hydrogen}\label{app:fcme_hy}
\begin{figure}[h!]
\includegraphics[width=0.5\textwidth]{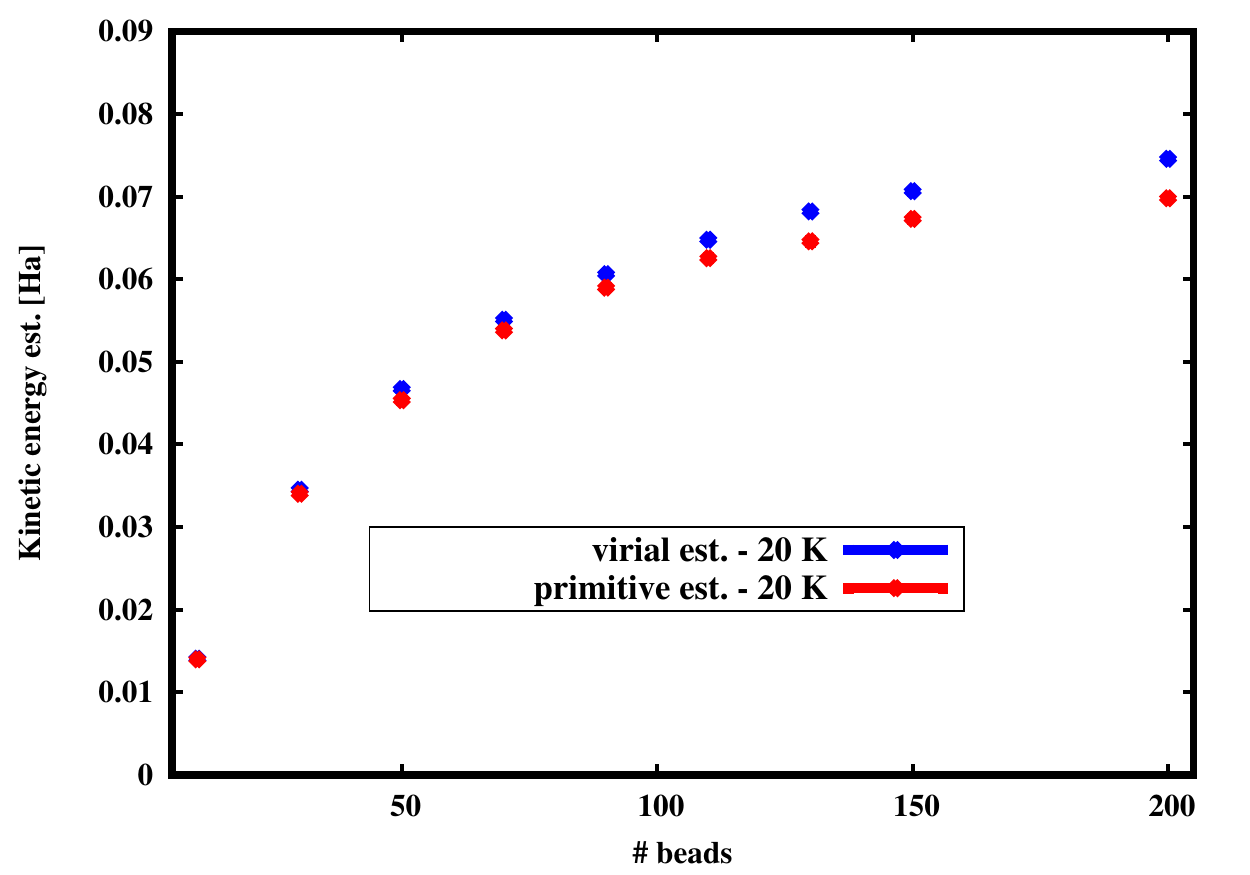}
\caption{\label{fig:hy_beads} Convergence of the virial (blue) and primitive (red) kinetic energy estimators at 20 K for the I4$_1$/amd atomic hydrogen.}
\end{figure}

\begin{figure}[h!]
\includegraphics[width=0.5\textwidth]{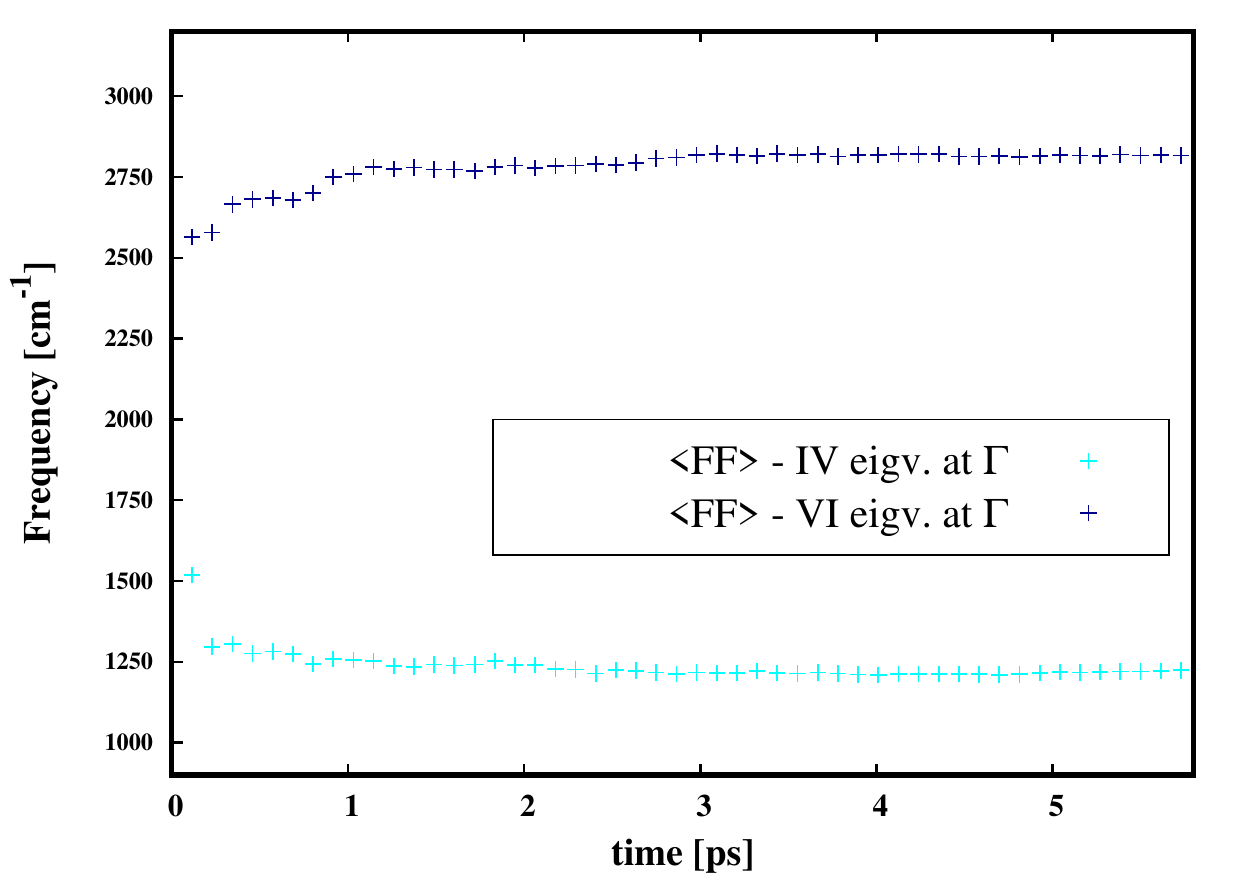}
\caption{\label{fig:hy_gamma} Convergence of the optical phonon frequencies at $\Gamma$ point for the $\langle FF \rangle$ estimator.}
\end{figure}

\begin{figure}[h!]
\includegraphics[width=0.5\textwidth]{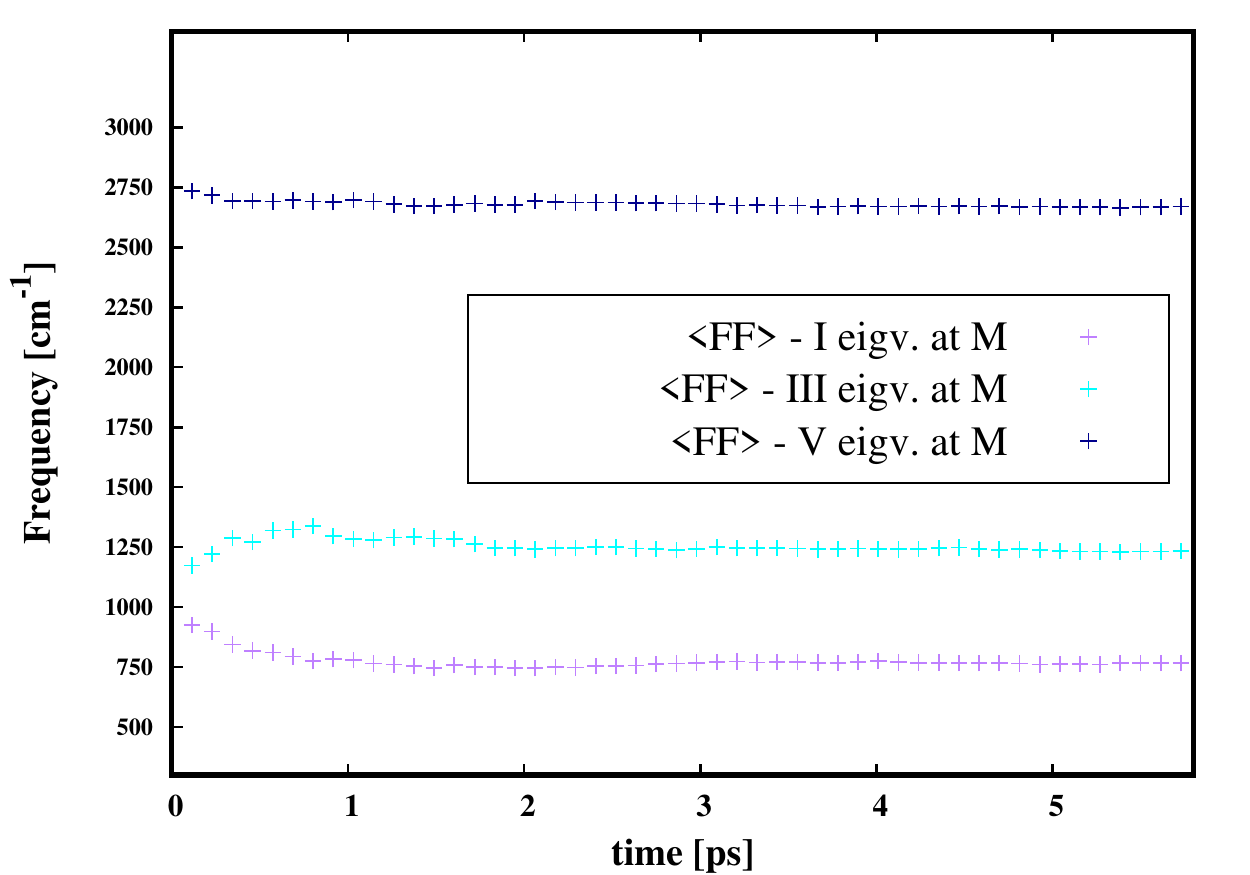}
\caption{\label{fig:hy_M} Convergence of phonon frequencies at the M point of the Brillouin zone for the $\langle FF \rangle$ estimator.}
\end{figure}

We present in Fig.~\ref{fig:hy_beads} the convergence of the kinetic energy estimators (virial and primitive \cite{Mouhat_2017}) for I4$_1$/amd hydrogen with respect the number of beads at 20 K. 
Although using 120 beads the kinetic energy is not fully converged, we choose this value for our simulations for computational reasons. Indeed, a larger value would have implied a lower time-step value and, of course, a longer simulation with additional CPU-time. The cost of this simulation was already around $2.5 \cdot 10^5$ total CPU-hours.\\ 
Furthermore, we show that the length of $\sim$6 ps is enough for the PIMD simulations of I4$_1$/amd hydrogen in order to get well converged phonon frequencies. As examples, in Fig.~\ref{fig:hy_gamma} we report the results at $\Gamma$, while in Fig.~\ref{fig:hy_M} the results at the M point. 
After 5 ps, we do not observe any significant change in the frequencies at the q-point sampled with the simulations and also in the phonon dispersions obtained from the successive interpolations. 

\nocite{*}
\bibliography{aipsamp}

\end{document}